\documentclass[twocolumn,aps,prd,longbibliography]{revtex4-1}
\usepackage{graphicx}  
\usepackage{dcolumn}   
\usepackage{bm}        
\usepackage{verbatim}   
\usepackage{graphicx}
\usepackage{epstopdf}
\usepackage{psfrag}
\usepackage{float}
\usepackage{footnote}
\usepackage{epsfig}
\usepackage{multirow}
\usepackage{subfigure}
\usepackage{amssymb}
\usepackage{xcolor}
\usepackage{amsmath}
\usepackage[most]{tcolorbox}
\usepackage{lipsum} 
\usepackage{color}
\usepackage[colorlinks=true,linkcolor=blue,allcolors=blue]{hyperref}%

\begin{document}
	
{PHYSICAL REVIEW APPLIED,~Vol.~**,  **~(2019)}
	
	\headsep = 40pt
	\title{Generalized Space-Time Periodic Diffraction Gratings:\\Theory and Applications}
	\author{Sajjad Taravati and George V. Eleftheriades}
	\affiliation{The Edward S. Rogers Sr. Department of Electrical and Computer Engineering, University of Toronto, Toronto, Ontario M5S 3H7, Canada\\
	Email: sajjad.taravati@utoronto.ca}

\begin{abstract}
This paper studies the theory and applications of the diffraction of electromagnetic waves by space-time periodic (STP) diffraction gratings. We show that, in contrast with conventional spatially periodic grating, a STP diffraction grating produces spatial diffraction orders, each of which is formed by an infinite set of temporal diffraction orders. Such spatiotemporally periodic gratings are endowed with enhanced functionalities and exotic characteristics, such as asymmetric diffraction pattern, nonreciprocal and asymmetric transmission and reflection, and an enhanced diffraction efficiency. The theory of the wave diffraction by STP gratings is formulated through satisfying the conservation of both momentum and energy, and rigorous Floquet mode analysis. Furthermore, the theoretical analysis of the structure is supported by time and frequency domain FDTD numerical simulations for both transmissive and reflective STP diffraction gratings. Additionally, we provide the conditions for Bragg and Raman-Nath diffraction regimes for STP gratings. Finally, as a particular example of a practical application of the STP diffraction gratings to communication systems, we propose an original multiple access communication system featuring full-duplex operation. STP diffraction gratings are expected to find exotic practical applications in communication systems, especially for the realization of enhanced-efficiency or full-duplex beam coders, nonreciprocal beam splitters, nonreciprocal and enhanced-resolution holograms, and illusion cloaks.
\end{abstract}

\maketitle		

\section{Introduction}

Light diffraction by spatially periodic structures is a fundamental phenomenon in optics and is of great importance in a variety of engineering applications~\cite{gaylord1982planar}. Such spatially periodic diffraction gratings are formed by a slab with a periodic \textit{spatial} variation at the \textit{wavelength scale}. The form of the grating periodicity is usually sinusoidal~\cite{tamir1964wave,burckhardt1966diffraction} or binary~\cite{moharam1995formulation}. They exhibit unique spectral properties
as the light impinging on the periodically modulated medium is reflected or transmitted at
specific angles only, which in general is not the case for aperiodic media. Diffraction gratings play the main role in numerous electromagnetic systems~\cite{hutley1982diffraction,loewen1997diffraction}, including but not restricted to, holography~\cite{newswanger1994holographic,smith2006holographic}, beam shaping~\cite{veldkamp1982laser}, data processing and opical logic~\cite{preston1972coherent,chavel1980optical}, medical diagnostic measurements~\cite{eichler1973thermal,phillion1975subnanosecond}, microwave and optical spectrum analysis~\cite{hecht1977spectrum,suhara1982folded,gaylord1985analysis}
		
Over the past decade, STP media have spurred a huge scientific attention, due to their extraordinary interaction with electromagnetic waves~\cite{cassedy1965waves,Cassedy_PIEEE_1967,Taravati_PRB_2017,Taravati_PRAp_2018,Taravati_Kishk_TAP_2019,inampudi2019rigorous,elnaggar2019generalized,Taravati_Kishk_PRB_2018,wang2018photonic,taravati2019space}. Such media are not governed by the Lorentz reciprocity law, so that they may provide a nonreciprocal frequency generation and amplification. Analytical investigation of wave propagation and scattering in time periodic media~\cite{zurita2009reflection,martinez2018parametric,salary2018time,wu2018transparent}, and STP media~\cite{Tien_JAP_1958,Oliner_PIEEE_1963,Fan_NPH_2009,Taravati_PRB_2017,Taravati_PRAp_2018,elnaggar2019generalized,li2019nonreciprocal,oudich2019space} represents an interesting topic due to the complexity and rich physics of the problem. Moreover, an interesting feature is the diverse and unique practical applications of STP media. As of today, STP structures have been used as parametric traveling-wave amplifiers~\cite{Cullen_NAT_1958,Tien_JAP_1958,tien1958traveling,Oliner_PIEEE_1963,li2019nonreciprocal}, optical isolators and circulators~\cite{wentz1966nonreciprocal,bhandare2005novel,Fan_PRL_109_2012,estep2014magnetic,Taravati_PRB_2017,Taravati_PRB_SB_2017,taravati2019space}, nonreciprocal metasurfaces~\cite{Alu_PRB_2015,Fan_APL_2016,Fan_mats_2017,Salary_2018,zhang2018space}, pure frequency mixers~\cite{Taravati_PRB_Mixer_2018}, antennas~\cite{shanks1961new,Taravati_APS_2015,Alu_PNAS_2016,ramaccia2018nonreciprocity,taravati2018space,salary2019nonreciprocal}, impedance matching structures~\cite{shlivinski2018beyond}, and mixer-duplexer-antenna systems~\cite{Taravati_LWA_2017}. 

The key contributions of this paper are as follows.
\begin{itemize}
\item Despite the recent surge of scientific interest on exploring outstanding and unique properties and applications of STP media~\cite{estep2014magnetic,Alu_PRB_2015,Taravati_APS_2015,Fan_APL_2016,Alu_PNAS_2016,Taravati_LWA_2017,Fan_mats_2017,Taravati_PRB_2017,Taravati_PRB_SB_2017,correas2018magnetic,Taravati_PRAp_2018,wu2018transparent,liu2018huygens,taravati2018advanced,Taravati_Kishk_PRB_2018,martinez2018parametric,Taravati_PRAp_2018,Salary_2018,Taravati_PRB_Mixer_2018,ramaccia2018nonreciprocity,taravati2018space,salary2019nonreciprocal,shlivinski2018beyond,li2019nonreciprocal,Taravati_Kishk_TAP_2019,inampudi2019rigorous,taravati2019space,elnaggar2019generalized}, there is still a lack of information on the operation of STP media in the \textit{diffraction regime}. Here, we first introduce the concept of \textit{generalized periodic gratings}. Such gratings are varying in both \textit{space and time}, representing the generalized version of standard conventional static (time-invariant) spatially varying gratings. Next, we provide a deep analysis on the functionality of STP gratings in the diffraction regime based on the \textit{modal analysis} for electromagnetic waves inside a STP grating and the \textit{wavevector-diagram analysis} for diffracted waves outside the STP grating.
\item We derive the electromagnetic wave diffraction from STP gratings and show that such gratings provide an infinite number of spatial diffraction orders, each of which composed of an infinite number of temporal diffraction orders. Such a unique spatial-temporal diffraction mechanism occurrs even upon incidence of a monochromatic wave on the STP grating.
\item It is shown that each single spatial-temporal diffraction order is diffracted at a distinct angle of diffraction, corresponding to a distinct wave amplitude. The provided general analysis is applicable to \textit{all types of periodicities}, e.g., binary and sawtooth periodic gratings. In addition, the presented \textit{generalized analytical solution} can be applied to spatial time-invariant gratings, spatially invariant temporal gratings, as well as spatial-temporal gratings.
\item It is demonstrated that, in contrast to conventional spatially periodic static gratings~\cite{moharam1983three,gaylord1985analysis,loewen1997diffraction,xu2015steering,bonod2016diffraction,memarian2012evanescent,memarian2013enhanced}, in a STP grating each spatial diffraction order is composed of an infinite number of temporal harmonics, each one of which is diffracted at a certain angle of diffraction.
\item We show that a STP grating provides various functionalities, such as nonreciprocal and angle (or system)-asymmetric wave diffraction, an asymmetric diffraction pattern and frequency conversion, which may be used to realize new optical and communication systems with enhanced efficiency.
\item The impact of the thickness of the STP grating on the diffraction mechanism is studied. It is shown that by varying the thickness of a STP grating, two completely different operation regimes, that is, Bragg (thick) and Raman-Nath (thin) regimes, can be achieved. The analytical formulas for the characteristics of these two regimes, and the efficiency and operation of these gratings are provided. 
\item We investigate the wave diffraction in both transmissive and reflective STP diffraction gratings. In particular, we present the asymmetric and nonreciprocal diffraction transmission in a transmitted STP grating and angle-asymmetric and nonreciprocal diffraction reflection in a reflective STP grating. We show that, in contrast to the transmissive STP grating, a reflective STP grating provides strong diffraction orders even if its thickness is subwavelength. This is due to the fact that, a reflective STP grating offers a much stronger interaction with the incident wave in comparison with transmissive STP gratings.
\item The provided theoretical analysis of general STP gratings is supported by FDTD numerical simulations. We present both time and frequency domains results, which provide a strong tool for investigation and understanding of the wave diffraction from general STP gratings.
\item We leverage some of the exotic properties and unique functionalities of STP diffraction gratings and present an advanced practical application. The proposed system is called space-time \textit{diffraction code multiple access} (STDCMA) system, which is an original multiple access communication system featuring \textit{full-duplex operation}.
\end{itemize}

The paper is structured as follows. Section~\ref{sec:theory} presents the theoretical analysis of the wave diffraction from general STP diffraction gratings, and derives the diffraction angles for each space-time diffracted order. In Sec.~\ref{sec:num}, we provide illustrative examples supported by the FDTD numerical simulation investigation in the time and frequency domains as follows: Section~\ref{sec:conv} investigates the wave diffraction from conventional static time-invariant gratings. Then, Sec.~\ref{sec:asym_patt} characterizes the STP grating, and shows its asymmetric diffraction pattern for normal incidence. Section~\ref{sec:asym_patt} also evaluates the effect of the grating thickness on the wave diffraction. Subsequently, Sec.~\ref{sec:nonr_asymm} demonstrates the nonreciprocal and asymmetric diffraction introduced by transmissive and reflective STP gratings. Section~\ref{sec:app} presents practical applications of STP gratings by leveraging the unique and exotic properties of their diffraction pattern. Finally, Sec.~\ref{sec:conc} concludes the paper.


\section{Theoretical Analysis}\label{sec:theory}

\subsection{Space-time periodic (STP) diffraction grating}
\begin{figure}
	\begin{center}
				\subfigure[]{\label{Fig:grating_conv}
			\includegraphics[width=0.95\columnwidth]{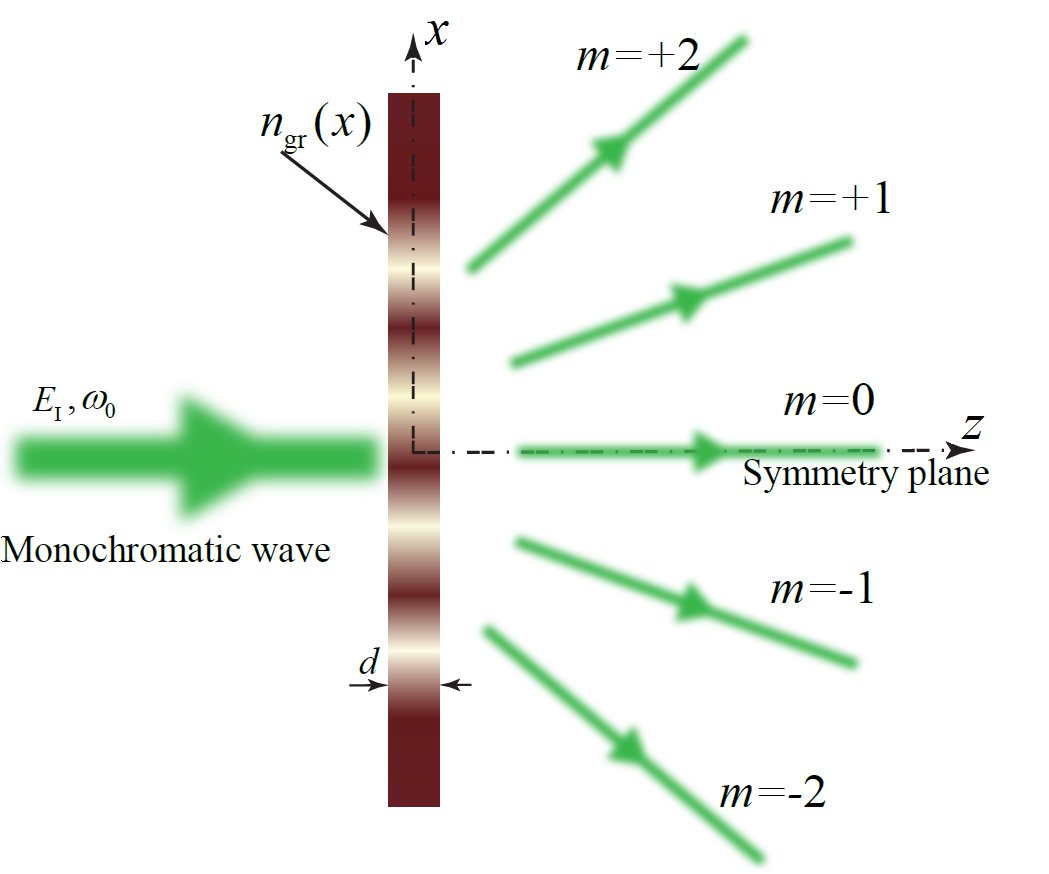}}
				\subfigure[]{\label{Fig:grating_ST}
			\includegraphics[width=1\columnwidth]{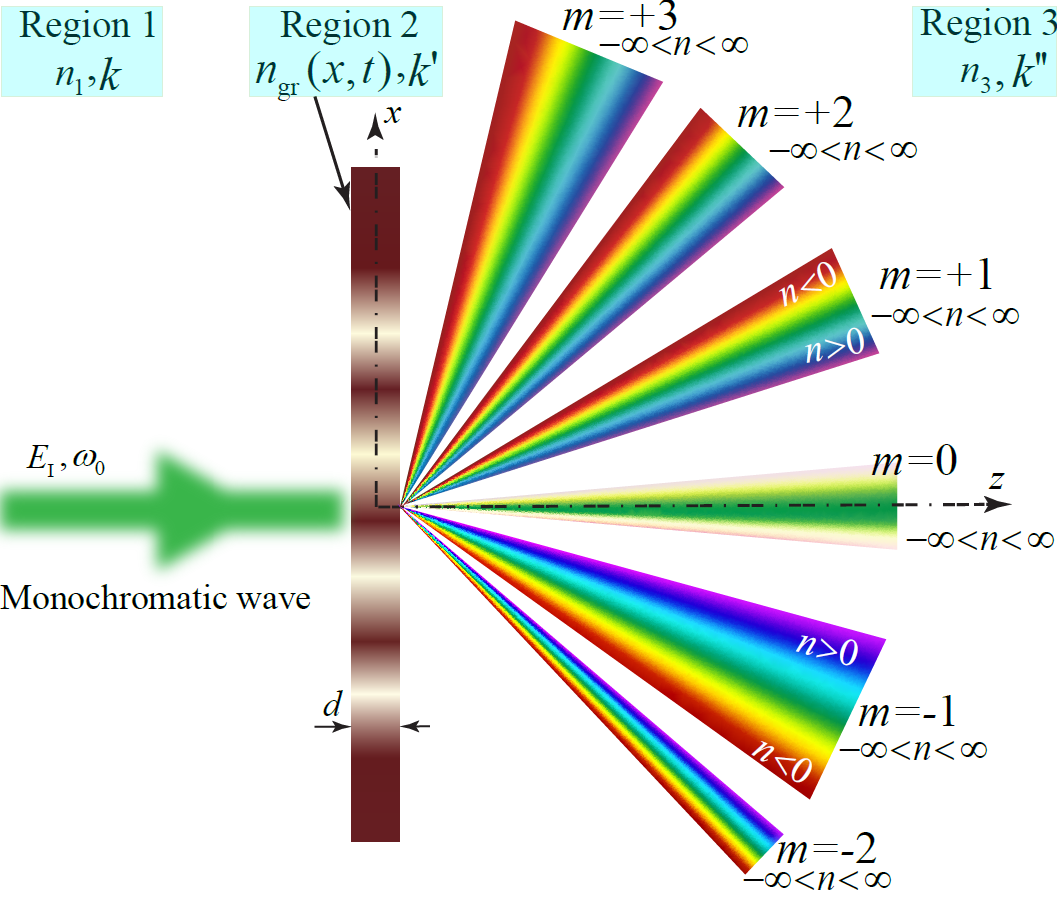}}
		\caption{Diffraction from a transmissive grating for a monochromatic incident wave. (a) Conventional spatial diffraction grating with $n_\text{gr}^2(x)= \epsilon_\text{gr}(x)=f_\text{per} (x)$, where spatial diffraction orders (e.g. $-2<m<2$) share the same temporal frequency, i.e., $\omega_0$, with the input wave.
		(b) Generalized STP diffraction grating, i.e., $n_\text{gr}^2(x,t)= \epsilon_\text{gr}(x,t)=f\left(f_\text{1,per} (x),f_\text{2,per} (t) \right)$, where each $m$th spatial diffraction order (e.g., $-2<m<3$) is formed by an infinite number of temporal diffraction orders $\omega_0+n\Omega$ with $-\infty<n<\infty$.}
		\label{Fig:sch}
	\end{center}
\end{figure}

Figure~\ref{Fig:grating_conv} depicts the wave diffraction from conventional transmissive planar spatially periodic diffraction gratings. The conventional static grating in Fig.~\ref{Fig:grating_conv} possesses a relative electric permittivity in the region from $z = 0$ to $z = d$ given by $n_\text{gr}^2(x)= \epsilon_\text{gr}(x)=f_\text{per} (x)$, where $f_\text{per} (x)$ is a periodic function of $x$, e.g., a sinusoidal, binary (square), or sawtooth function. Electromagnetic waves always travel in straight lines, but when passing near an obstruction they tend to bend around that obstruction and spread out. The diffraction phenomenon occurs when an electromagnetic wave passes by a corner or through a slit or grating that has an optical size comparable to the wavelength. The diffraction by a grating is a specialized case of wave scattering, where an object with regularly repeating features yields an orderly diffraction of the electromagnetic wave in a pattern consisting of a set of diffraction orders $m$.

As shown in Fig.~\ref{Fig:grating_conv}, considering normal incidence of the input wave ($\theta_\text{i}=0$), a symmetric diffraction pattern with respect to $x=0$ will be produced by conventional static gratings, possessing a symmetric profile with respect to the $x=0$ axis. An asymmetric diffraction pattern for normal incidence can be achieved by asymmetric static periodic metagratings~\cite{popov2018controlling,popov2019constructing}. However, gratings with symmetric and asymmetric profiles are both restricted by the Lortentz reciprocity theorem, and therefore, possess reciprocal diffraction transmission response. The symemtry of the diffraction pattern in conventional periodic static gratings includes the symmetry in both the angles of diffraction orders $\theta_{m}$ (e.g., $\theta_{+2}=\theta_{-2}$) and the symmetry in the intensity of the diffracted orders $P_m$  (e.g., $P_{+2}=P_{-2}$). In addition, assuming a monochromatic input wave with temporal frequency $\omega_0$, no change in the temporal frequency of the incident field occurs, and hence, the diffracted orders share the same temporal frequency of $\omega_0$.
	
Now, consider the transmissive planar STP diffraction grating shown in Fig.~\ref{Fig:grating_ST}. This figure shows a generic representation of the spatiotemporal diffraction from a STP diffraction grating, which is distinctly different from the spatial diffraction from a conventional space periodic diffraction grating in Fig.~\ref{Fig:grating_conv}. The grating is interfaced with two semi-infinite dielectric regions, i.e., region 1 characterized with the refractive index $n_1$ and wavenumber $k$, and region 3 characterized with the refractive index $n_3$ and wavenumber $k''$. The relative electric permittivity of this STP grating is periodic in both space and time, with temporal frequency $\Omega$ and spatial frequency $K$, given by
\begin{equation}\label{eqa:eps}
n_\text{gr}^2(x,t)= \epsilon_\text{gr}(x,t)=f\left(f_\text{1,per} (x),f_\text{2,per} (t) \right),
\end{equation} 
where $f_\text{1,per} (x)$ and $f_\text{2,per} (t)$ are periodic functions of space (in the $x$ direction) and time, respectively. The wavenumber in region 2 (inside the STP grating) is denoted by $k'$.

Assuming normal (or oblique) incidence of the input wave, the spatiotemporally periodic gratings (shown in Fig.~\ref{Fig:grating_ST}) produces an asymmetric diffraction pattern with respect to $x=0$. This asymmetry in the diffraction pattern is due to the asymmetric spatiotemporal profile of the structure provided by the space-time modulation. The asymmetry of the diffraction pattern extends to both the diffraction angles of diffracted orders $\theta_{m}$ (e.g., $\theta_{m=+2} \neq \theta_{m-2}$) and the intensities of the diffracted orders $P_m$  (e.g., $P_{m=+2} \neq P_{m=-2}$). Furthermore, the time-variation of the grating (with frequency $\Omega$) results in the generation of new frequencies. Hence, assuming a monochromatic input wave with temporal frequency $\omega_0$, an infinite set of temporal frequencies will be generated inside the grating and will be diffracted, so that each spatial diffracted order ($m$) is composed of an infinite number of temporal diffraction orders $n$. As a result, for such a generalized STP diffraction grating, the diffraction characteristics are defined for each spatial-temporal diffracted order ($mn$) so that the diffracted order ($mn$) is transmitted at a specified angle $\theta_{mn}$ attributed to the electric field $E_{mn}^\text{T}$.

\begin{figure}
	\begin{center}
		\subfigure[]{\label{Fig:boundary_sp}
			\includegraphics[width=0.48\columnwidth]{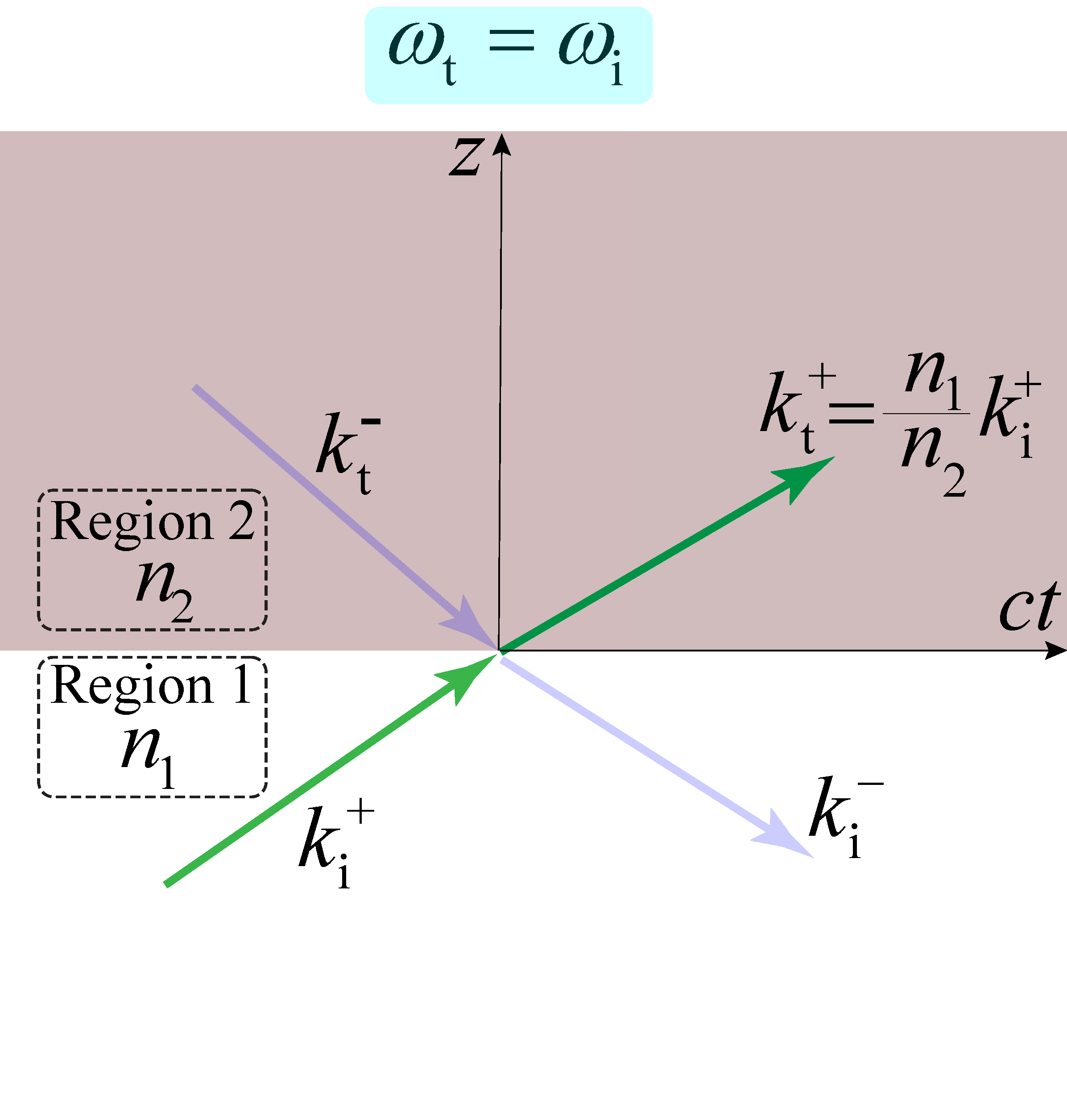}}
					\subfigure[]{\label{Fig:boundary_temp}
				\includegraphics[width=0.48\columnwidth]{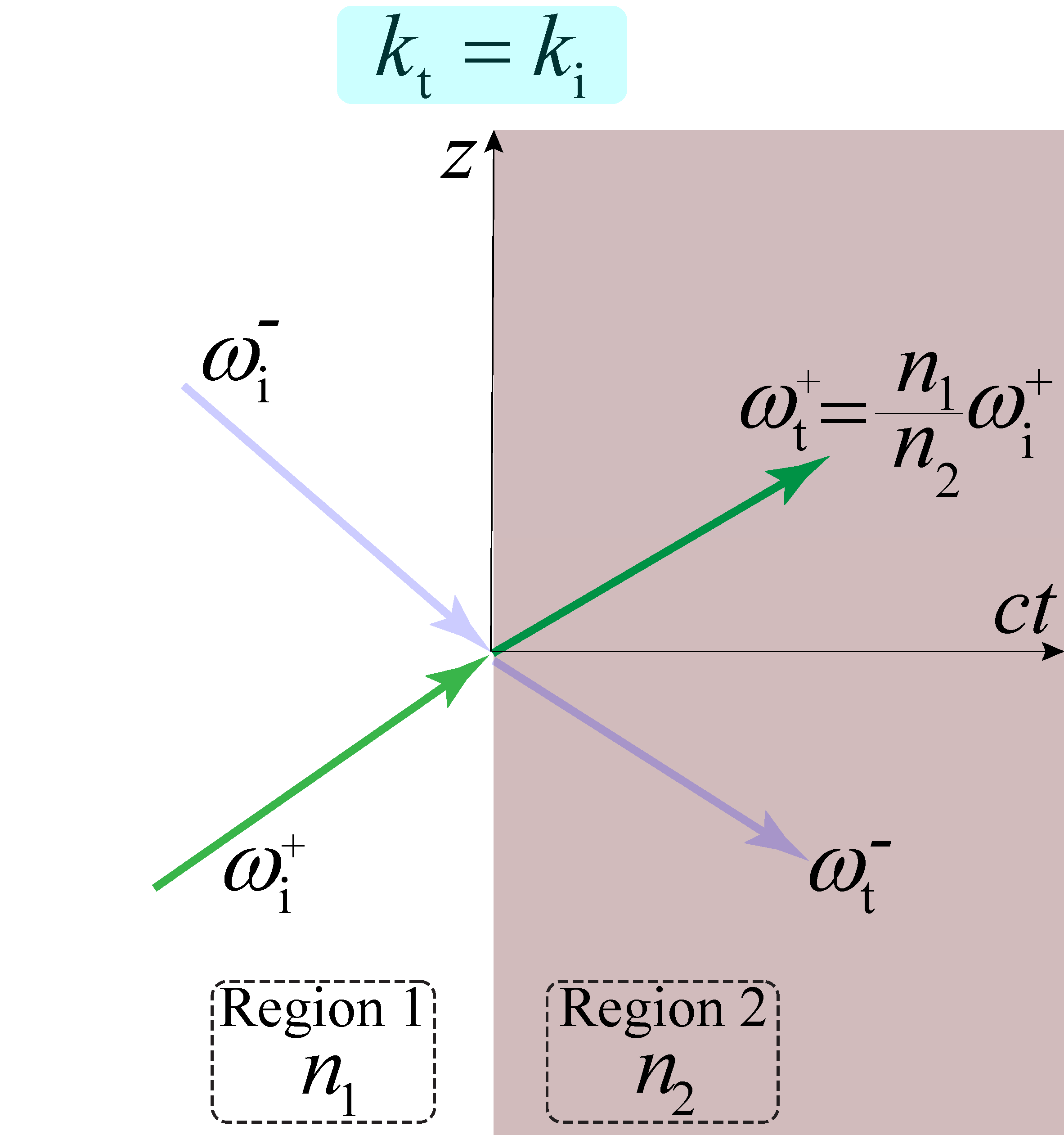}}
		\subfigure[]{\label{Fig:boundary_ST}
			\includegraphics[width=0.7\columnwidth]{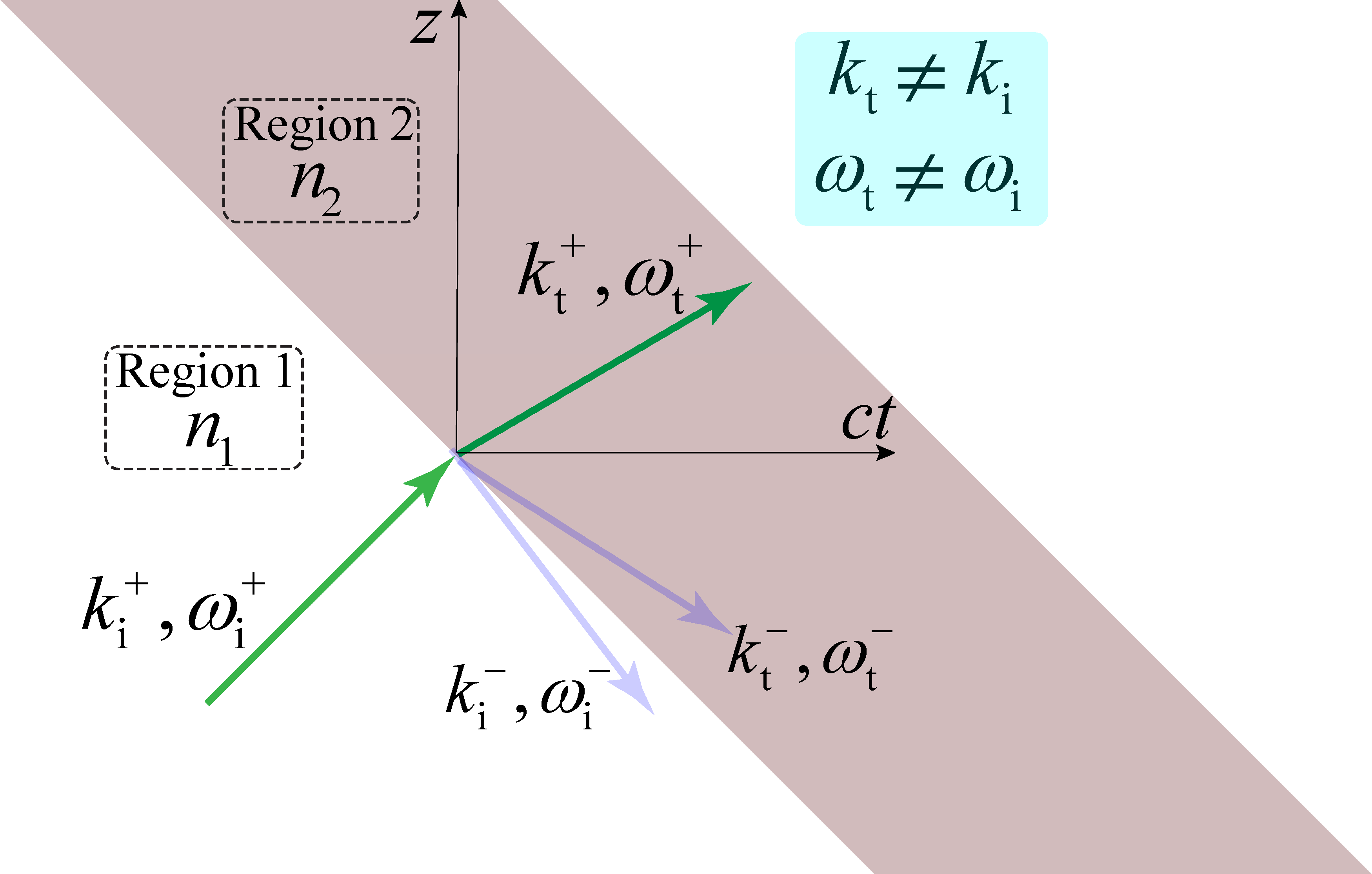}}
		\caption{Space-time diagrams showing scattering of forward and backward fields 
		and conservation of energy and momentum for different scenarios. (a) Spatial interface, i.e., $n(z<0)=n_1$ and $n(z>0)=n_2$. (b) Temporal interface, i.e., $n(t<0)=n_1$ and $n(t>0)=n_2$. (c) Spatial-temporal interface, i.e., $n(z/c+t<0)=n_1$ and $n(z/c+t>0)=n_2$.}
		\label{Fig:boundary}
	\end{center}
\end{figure}

To best investigate the wave diffraction by a space-time-varying grating, we first study the interaction of the electromagnetic wave with space and time interfaces, separately. Figure~\ref{Fig:boundary_sp} sketches the Minkowski space-time diagram of a spatial interface between two media of refractive indices $n_1$ and $n_2$, respectively, in the plane ($z,ct$). This figure shows scattering of forward and backward fields and conservation of energy and momentum for different scenarios. The temporal axis of the Minkowski space-time diagram is scaled with the speed of light $c$, and therefore is labeled by $ct$ for changing the dimension of the addressed physical quantity from time to length, in accordance to the dimension associated to the spatial axes labeled $z$. This problem represents the conventional case of electromagnetic wave incidence and scattering from a spatial (static) interface, where $n(z<0)=n_1$ and $n(z>0)=n_2$. At this spatial boundary, the normal magnetic field $\textbf{B}$, the normal electric field displacement $\textbf{D}$, and the temporal frequency are preserved, but the wavenumber $k$ changes, i.e., energy is preserved but momentum changes. As a result, the wavenumber of the forward transmitted wave in the region 2 corresponds to $k^+_\text{t}=n_1 k^+_\text{i}/n_2$, whereas the temporal frequency of the transmitted wave in region 2 is equal to that of the region 1, i.e., $\omega_\text{t}=\omega_\text{i}$.

Figure~\ref{Fig:boundary_temp} shows the space-time diagram of a time interface between two media of refractive indices $n_1$ and $n_2$, which is the dual case of the spatial slab in Fig.~\ref{Fig:boundary_sp}~\cite{felsen1970wave,biancalana2007dynamics,bacot2016time}. Here, the refractive index suddenly changes from one value ($n_1$) to another ($n_2$) at a given time throughout all space, i.e., $n(t<0)=n_1$ and $n(t>0)=n_2$. The temporal change of the refractive index produces both reflected (backward) and transmitted (forward) waves, which is analogous to the reflected and transmitted waves
produced at the spatial interface between two different media in Fig.~\ref{Fig:boundary_sp}. The total charge $Q$ and the total flux $\psi$ must remain constant at the moment of the jump from $n_1$ to $n_2$, implying that both transversal and normal components of $\textbf{D}$ and $\textbf{B}$ do not change instantaneously~\cite{morgenthaler1958velocity,fante1971transmission}, which is different than the static case (shown in Fig.~\ref{Fig:boundary_sp}) where only normal components of the magnetic field $\textbf{B}$ and electric field displacement $\textbf{D}$ are conserved. Specifically, at a time interface, the magnetic field $\textbf{B}$, the electric field displacement $\textbf{D}$ and the wavenumber $k$ are preserved. This yields a change in the temporal frequency of the incident wave so that the frequency of the forward transmitted wave in the region 2 corresponds to $\omega^+_\text{t}=n_1 \omega^+_\text{i}/n_2$, i.e., where momentum is preserved but energy changes.

Figure~\ref{Fig:boundary_ST} depicts the space-time diagram of a spatial-temporal interface, i.e., $n(z/c+t<0)=n_1$ and $n(z/c+t>0)=n_2$, as the combination of the space and time interfaces in Figs.~\ref{Fig:boundary_sp} and~\ref{Fig:boundary_temp}, respectively.
It may be seen that the spatial-temporal interface resembles the spatial interface configuration in Fig.~\ref{Fig:boundary_sp} in the region $n=n_1$ and the temporal interface configuration in Fig.~\ref{Fig:boundary_temp} for $n=n_2$. Here, only one of the four forward and backward waves reaches the interaction point from the past, whereas the other three waves travel from the interface in the positive time direction~\cite{biancalana2007dynamics,deck2017scattering}. At such a spatial-temporal interface, normal component of the magnetic field $\textbf{B}$ and normal component of the electric field displacement $\textbf{D}$ are preserved~\cite{costen1965three,bolotovskiui1989reflection}. However, both the spatial frequency (wavenumber) $k$ and the temporal frequency changes, i.e., both momentum and energy change. For a periodic space-time-modulated medium, as in Fig.~\ref{Fig:grating_ST}, the same phenomenon, i.e., a change in the spatial and temporal frequencies occurs for each interface. Hence, following the Floquet theorem, a STP grating introduces an infinite number of space and time diffraction orders, as it is described in Sec.~\ref{sec:angles}.

\subsection{Diffraction angles}\label{sec:angles}
\begin{figure*}
	\includegraphics[width=2\columnwidth]{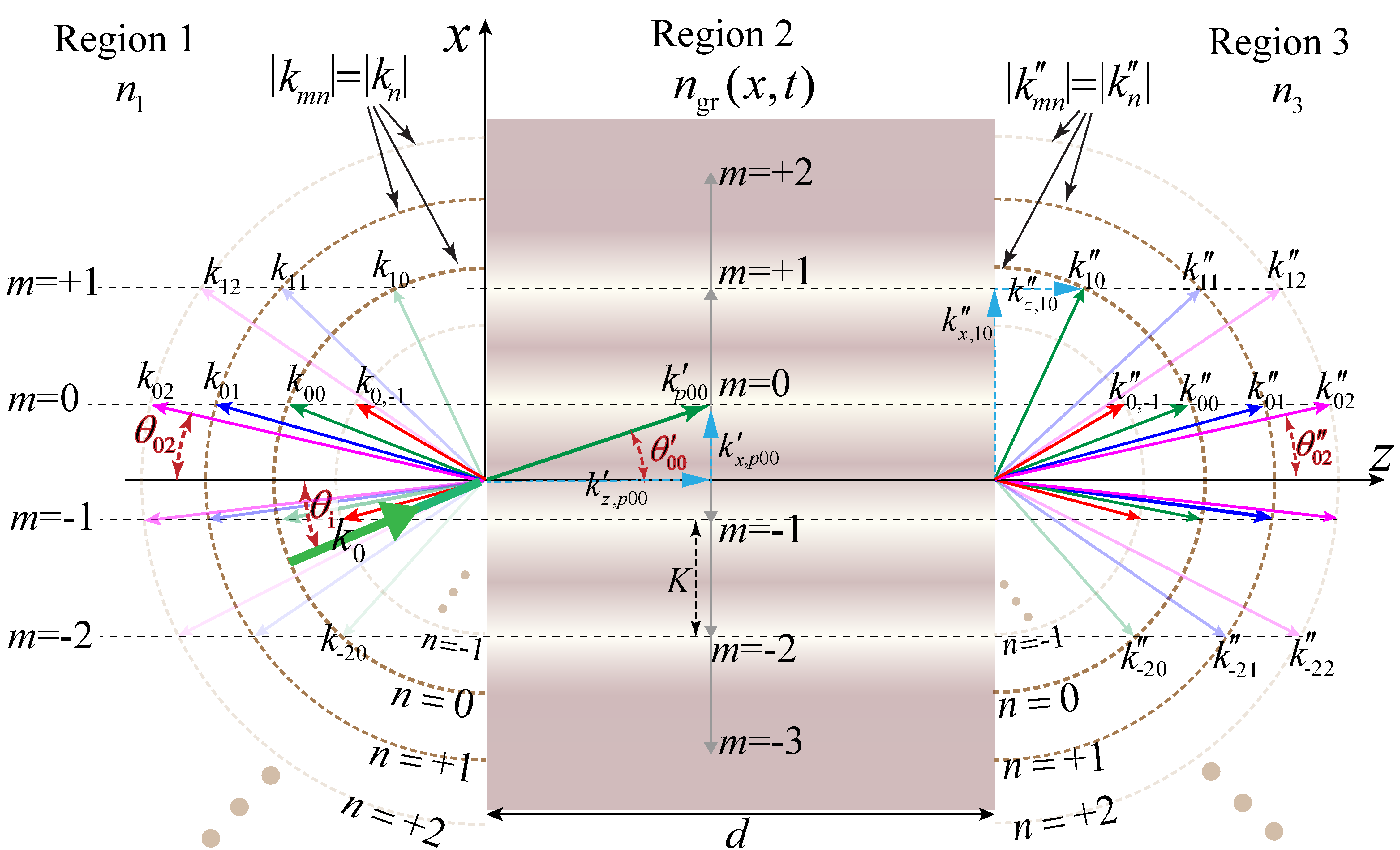}
	\caption{Wavevector isofrequency diagram for the diffraction from a STP diffraction grating, exhibiting phase matching of spatial-temporal harmonic components of total field inside the grating with propagating backward diffracted orders in region 1, and
		forward diffracted orders in region 3. The diffracted spatial-temporal harmonics corresponding to $-1 < m < +2$ are propagating diffracted orders, whereas the harmonics corresponding to $m<-1$ and $m>+2$ are evanescent (cut off) outside the STP grating.}
	\label{fig:analysis}
\end{figure*}

Figure~\ref{fig:analysis} shows a generic illustration example of a wavevector isofrequency diagram for the diffraction from a STP diffraction grating. The grating is characterized with the spatial frequency $K$ (the spatial periodicity of the STP grating reads $\Lambda=2\pi/K$) and the temporal frequency $\Omega$. Figure~\ref{fig:analysis} sketches the phase matching of spatial-temporal harmonic components of the total field inside the grating with propagating backward diffracted orders in region 1, and
forward diffracted orders in region 3. We assume the grating is interfaced with two semifinite dielectrics, i.e., $z\rightarrow -\infty<\text{region 1}<z=0$ and $d<\text{region 3}<z \rightarrow \infty$, respectively. Region 1, region 2 (inside the STP grating) and region 3 are, respectively, characterized with the phase velocities $v_\text{r}=c/n_1$, $v'_\text{r}=c/n_\text{av}$ and $v''_\text{r}=c/n_3$, and the wavevectors $\text{\textbf{k}}_{mn}= k_{x,mn} \mathbf{\hat{x}}+k_{z,mn} \mathbf{\hat{z}}$, $\text{\textbf{k}}'_{pmn}= k'_{x,pmn} \mathbf{\hat{x}}+k'_{z,pmn} \mathbf{\hat{z}}$, and $\text{\textbf{k}}''_{mn}= k''_{x,mn} \mathbf{\hat{x}}+k''_{z,mn} \mathbf{\hat{z}}$. Here, $c$ represents the velocity of the light in the vacuum, $m$ and $n$ denote the number of the space and time harmonics, respectively, while $p$ represents the number of the mode in region 2, inside the grating (these modes only exist inside the grating, and Sec.~\ref{eqa:fields} elaborates on their properties). 	
	
The STP grating assumes oblique incidence of a $y$-polarized electric field, with temporal frequency $\omega_0$ and under an angle of incidence $\theta_{\text{i}}$ with respect to the normal of the grating, i.e.,
\begin{equation}\label{eqa:Ei}
\mathbf{E}_\text{i} (x,z,t)= \mathbf{\hat{y}} E_0 e^{i\left( k_{x,\text{i}} x +k_{z,\text{i}} z- \omega _0 t  \right) },
\end{equation}
where $E_0$ is the amplitude of the incident wave. In Eq.~\eqref{eqa:Ei}, $k_{x,\text{i}}=k_{0} \sin(\theta_{\text{i}})=\omega_{0} \sin(\theta_{\text{i}})/v_\text{r}$ and $k_{z,\text{i}}=k_{0} \cos(\theta_{\text{i}})$ are the $x$ and $z$ components of the incident wavevector, respectively. 

The $x$ component of the wavevector outside the STP grating, in region 3, reads
\begin{equation}\label{eq:kx}
k''_{x,mn}=k''_n \sin(\theta''_{mn}),
\end{equation}
where 
\begin{equation}\label{eq:kx}
k''_n =k''_0+n \frac{\Omega}{v''_\text{r}},
\end{equation}
and where $k''_0=\omega_0/v''_\text{r}$. The corresponding $z$ component of the wavevector in region 3 is calculated using the Helmholtz relation, as
\begin{equation}\label{eq:kz}
k''_{z,mn}=\sqrt{(k''_{mn})^2-(k''_{x,mn})^2}=k''_n \cos(\theta''_{mn}).
\end{equation} 

The $x$ and $z$ components of the wavevector in region 1 ($k_{x,mn}$ and $k_{z,mn}$) and inside the grating ($k'_{x,mn}$ and $k'_{z,mn}$) can be achieved following the same procedure as in Eqs.~\eqref{eq:kx} and~\eqref{eq:kz}. The space-time diffraction process may be simply interpreted as follows. The incident wave is refracted into the grating medium at $z = 0$, while generating an infinite set of time harmonics inside the grating, with frequencies $\omega_n=\omega_0+n \Omega$ corresponding to the wavevectors $k'_n=k'_0+n \Omega/v'_\text{r}$. The refracted space-time plane waves in the grating are diffracted into an infinite set of plane waves traveling toward the $z = d$ boundary. The space-time harmonic waves inside the grating are phase matched into propagating and evanescent waves in region 3, i.e., the $x$ components of the wavevectors of the $m$th mode in regions 1 and 3 and the $x$ component of the wavevector of the $m$th space-time harmonic field in region 2 must be the same.

To determine the spatial and temporal frequencies of the diffracted orders, we consider the momentum conservation law, i.e.,
\begin{subequations}
\begin{equation}\label{eqa:gr_eq01}
k_{x,\text{diff}}=k_{x,\text{inc}}+ m K,
\end{equation} 
or
\begin{equation}\label{eqa:gr_eq02}
k''_{x,mn}=k_{x,mn}=k_{x,\text{i}}+ mK,
\end{equation} 
and the energy conservation law, i.e.,
\begin{equation}\label{eqa:gr_eq13}
\omega_\text{diff}=\omega_\text{inc}+ n \Omega,
\end{equation} 
or
\begin{equation}\label{eqa:gr_eq14}
\omega_n=\omega_0+ n \Omega,
\end{equation} 
\end{subequations}
where $k_{x,\text{diff}}$ and $k_{x,\text{inc}}$ denote the $x$ components of the wavevector of the diffracted and incident fields, respectively, and $\omega_\text{diff}$ and $\omega_\text{inc}$ represent the temporal frequencies of the diffracted and incident fields, respectively. Equation~\eqref{eqa:gr_eq02}, using~\eqref{eq:kx}, may be written as
\begin{equation}
\left(k''_0+n \frac{\Omega}{v''_\text{r}} \right) \sin\left(\theta''_{mn} \right)=k_0 \sin(\theta_\text{i})+mK,
\end{equation} 
where $k_0=n_1\omega_0/c$. Seeking for the angle of diffraction for the forward spatial-temporal diffracted orders in region 3, i.e., the $m$th spatial and $n$th temporal harmonic, yields	

\begin{equation}\label{eqa:refl_trans_angl}
\sin\left(\theta''_{mn} \right)=\frac{n_1}{n_3} \frac{\sin(\theta_\text{i})+m K/k_0}{1+n \Omega/\omega_0}   ,
\end{equation}

The corresponding angle of diffraction for the backward diffracted orders in region 1 reads
\begin{equation}\label{eqa:refl_angl}
\sin\left(\theta_{mn} \right)=\frac{\sin(\theta_\text{i})+m K/k_0}{1+n \Omega/\omega_0}  
\end{equation}	

\subsection{Propagating and evanescent Orders}
For a given set of incident angles, spatial and temporal frequencies of the grating, and the wavelength of the incident beam, the grating equation in Eq.~\eqref{eqa:gr_eq02} may be satisfied for more than one value of $m$ and $n$. However, there exists a solution only when $|\sin\left(\theta_{mn} \right)|<1$. Diffraction orders corresponding to $m$ and $n$ satisfying 
this condition are called \textit{propagating} orders. The other orders yielding $|\sin\left(\theta_{mn} \right)|>1$ correspond to imaginary $z$ components of the wavevector $k_{z,mn}$ as well as complex angles of diffraction $\sin(\theta_{mn})$. 
These evanescent orders decrease exponentially with the distance from the grating, and hence, can be detected only at a distance less than a few wavelengths from the grating. However, these evanescent orders play a key role in some surface-enhanced grating properties and are taken into account in the theory of gratings. Evanescent orders
are essential in some special applications, such as for instance waveguide and fiber gratings. The specular order ($m = 0$) is always propagating while the others can be either propagating or evanescent. The modulations with $2\pi/K<< \lambda_0$ will produce evanescent orders for $m \neq 0$, while the modulations with $2\pi/K>> \lambda_0$ will yield a large number of propagating orders.

In the homogeneous regions, i.e., regions 1 and 3, the magnitude of the wavevectors of the backward- and forward-diffracted orders read
\begin{equation}
|k_{mn}|=|k_n|, \qquad \text{and}  \qquad |k''_{mn}|=|k''_n|
\end{equation}
	
As explained before, the $x$ components of the diffracted wavevectors, $k_{x,mn}$ and $k''_{x,mn}$, can be deduced from the phase-matching requirements. Then, the propagating and evanescent nature of the corresponding orders will be specified based on the $k_{z,mn}$ and $k''_{x,mn}$, as follows. The real $k_{z,mn}$s and $k''_{z,mn}$s correspond to propagating orders, whereas the imaginary $k_{z,mn}$s and $k''_{z,mn}$s correspond to evanescent orders. The propagating and evanescent $m$th fields in
regions 1 and 3 are shown in Fig.~\ref{fig:analysis}. The wavevectors in regions 1 and 3 possess magnitudes $|k_{n}|$ and $|k''_n|$, respectively. Hence, all the spatial diffraction orders for the $n$th temporal harmonic in these two regions share the same amplitude, i.e, $|k_{mn}|=|k_n|$ and $|k''_{mn}|=|k''_n|$. Semicircles with these radii are sketched in Fig.~\ref{fig:analysis}. The allowed wavevectors in these regions must be phased matched to the boundary components of the spatial-temporal diffracted order inside the grating. This is shown by the horizontal dashed lines in the figure. In the qualitative illustration in Fig.~\ref{fig:analysis}, for the incident wave of wavevector $k_0$ and the grating with grating wavevector $K$ and temporal frequency $\Omega$, the $m = -1$ to $+2$ waves exist as propagating diffracted orders in regions 1 and 3. However, $m\leqslant-2$ and $m \geqslant +1$ will be diffracted as evanescent orders.

\subsection{Diffracted Electromagnetic fields}\label{eqa:fields}
The electromagnetic wave propagation and diffraction in general periodic media may be studied by several approaches. Among the proposed approaches, the modal approach~\cite{cassedy1963dispersion,tamir1964wave,gaylord1985analysis,Taravati_PRB_2017} and the coupled-wave approach~\cite{moharam1995formulation,Fan_NPH_2009} represent the most common and insightful approaches for analysis of periodic media diffraction gratings, both of which provide exact formulations without approximations. Here, we study the wave diffraction inside the STP grating using the modal approach. The modal approach has also been referred to as the Bloch-Floquet (or Floquet-Bloch), characteristic-mode, and eigenmode approach. Such an approach expresses the electromagnetic fields inside the grating as a combination of an infinite number of modes, each of those individually satisfying Maxwell’s equations. 

First, we expand the field inside the modulated medium in terms of the spatial-temporal diffracted orders ($m$ and $n$) of the field in the periodic structure. This is due to the fact that the electromagnetic waves in periodic media take on the same periodicity as their host. These spatial-temporal diffracted orders inside the grating are phase matched to diffracted orders (either propagating or evanescent) outside of the grating. The partial space-time harmonic fields may be considered as inhomogeneous plane waves with a varying amplitude along the planar phase front. These inhomogeneous plane waves are dependent and they exchange energy back and forth between each other in the modulated grating. 

Since the electric permittivity of the grating is periodic in both space and time, with spatial frequency $K$ and temporal frequency $\Omega$, it may be expressed in terms of the double Fourier series expansion, as
\begin{equation}\label{eqa:index}
n_\text{gr}^2(x,t)= \epsilon_\text{gr}(x,t)=\sum_{m}\sum_{n}  \epsilon_{mn} e^{i (mK x  -n \Omega t)} ,
\end{equation} 
where $\epsilon_{mn}$ are complex coefficients of the permittivity, and $K$ and $\Omega$ are the spatial and modulation frequencies, respectively. The electric field inside the grating is expressed in terms of a sum of an infinite number of modes, i.e.,
\begin{equation}\label{eqa:Em}
\begin{split}
\mathbf{E}_{2}(x,z,t)=  \sum_{p=-\infty}^{\infty} \mathbf{E}_{2,p}(x,z,t),
\end{split}
\end{equation} 

Given the spatial-temporal periodicity of the grating, the corresponding electric field of the $p$th mode inside the grating may be decomposed into spatiotemporal Bloch-Floquet plane waves, as
\begin{equation}\label{eqa:Emp}
\begin{split}
\mathbf{E}_{2,p}(x,z,t)=\mathbf{\hat{y}}  \sum_{m} \sum_{n} E'_{pmn} e^{i ( k'_{x,pmn} x +k'_{z,pmn} z -\omega_{n} t)} ,
\end{split}
\end{equation} 
where 
\begin{equation}\label{eqa:kka}
\begin{split}
k'_{x,pmn}=k'_{x,p0n}+mK&=\left(k'_{p00}+n \frac{\Omega}{v'_\text{r}} \right) \sin\left(\theta'_{p0n}\right)+ mK 
\end{split}
\end{equation} 
and
\begin{equation}\label{eqa:kkb}
\begin{split}
k'_{z,pmn}=k'_{pmn} \cos(\theta_{\text{i}}),
\end{split}
\end{equation} 

In Eq.~\eqref{eqa:kka}, $\theta'_{p0n}$ reads
\begin{equation}\label{eqa:kk2}
\theta'_{p0n} =\tan^{-1}\left( \frac{k'_{x,p0n}}{k'_{z,p0n}}  \right)
\end{equation} 

The corresponding magnetic field inside the grating reads
\begin{equation}\label{eqa:Hm}
\begin{split}
&\mathbf{H}_2 (x,z,t)=\dfrac{1}{\eta}  \mathbf{\hat{k}'}_{pmn} \times\mathbf{E}_2 (x,z,t)\\
&=\sum_{p,m,n}  \left(- \mathbf{\hat{x}} \frac{k'_{z,pmn}}{k'_{pmn}}  +\mathbf{\hat{z}} \frac{k'_{x,pmn}}{k'_{pmn}}  \right)
\frac{E'_{pmn}}{\eta' }  e^{i ( k'_{x,pmn} x +k'_{z,pmn} z -\omega_{n} t)} 
\end{split}
\end{equation} 

The unknown field coefficients $E'_{pmn}$ and $k'_{x,p00}$ are to be found through satisfying Maxwell's equations, that is,
\begin{subequations}
	\begin{equation}\label{eqa:Max1}
	\nabla\times\textbf{E}_\text{2} (x,z,t)=-\dfrac{\partial  \textbf{B}_\text{2} (x,z,t)}{\partial t}
	\end{equation}
	\begin{equation}\label{eqa:Max2}
	\nabla\times\textbf{H}_\text{2} (x,z,t)=\dfrac{\partial \textbf{D}_\text{2} (x,z,t)}{\partial t} 
	\end{equation}
\end{subequations}

The corresponding wave equation for the STP grating may be derived from Eqs.~\eqref{eqa:Max1} and~\eqref{eqa:Max2} and reads
\begin{equation}\label{eqa:Max3}
\nabla^2\textbf{E}_\text{2} (x,z,t)=\frac{1}{c^2} \dfrac{\partial^2}{\partial t^2}  \left[ \epsilon_\text{gr}(x,t) \textbf{E}_\text{2} (x,z,t) \right]
\end{equation}

We assume that the grating is invariant in the $y$ direction (i.e., $\partial/\partial y=0$). Then, inserting~\eqref{eqa:Em} into~\eqref{eqa:Max3} yields

\begin{equation}\label{eqa:Max2_b}
\begin{split}
&\left(\dfrac{\partial^2 }{\partial x^2} +\dfrac{\partial^2 }{\partial z^2} \right) E'_{pmn} e^{i ( k'_{x,pmn} x +k'_{z,pmn} z -\omega_{n} t)}\\
&=\frac{1}{c^2} \dfrac{\partial^2}{\partial t^2}  \left( \sum_{j}\sum_{q} \epsilon_{jq} E'_{pmn} e^{i ( [k'_{x,pmn}+jK] x +k'_{z,pmn} z -[\omega_{n}+q \Omega] t)} \right)\\
&=\frac{1}{c^2} \dfrac{\partial^2}{\partial t^2}   \sum_{j}\sum_{q} \epsilon_{m-j,n-q}  E'_{pjq} e^{i ( k'_{x,pmn} x +k'_{z,pmn} z -\omega_{n} t)} 
\end{split}
\end{equation}

Solving Eq.~\ref{eqa:Max2_b} for the unknown field coefficients $E'_{pmn}$ gives
\begin{equation}\label{eqa:Max2_c}
\begin{split}
& E'_{pmn}=\frac{(\omega_{n}/c)^2}{(k'_{x,pmn})^2 +(k'_{z,pmn})^2 }  \sum_{j}\sum_{q} \epsilon_{m-j,n-q}  E'_{pjq} 
\end{split}
\end{equation}

Next, we determine the backward diffracted fields in region 1 and forward diffracted fields in region 3. As depicted in Fig.~\ref{fig:analysis}, one must consider the multiple backward and forward-propagating diffracted orders that exist inside and outside of the grating. The total electric field in region 1 is the sum of the incident and the backward-traveling diffracted orders, as
\begin{equation}\label{eqa:ER}
\begin{split}
\mathbf{E}_\text{1} =\mathbf{\hat{y}} E_{0} e^{i ( k_{x,\text{i}} x +k_{z,\text{i}} z -\omega_{0} t)}+ \mathbf{\hat{y}} \sum_{m,n} E^\text{R}_{mn} e^{i ( k_{x,mn} x -k_{z,mn} z -\omega_{n} t)},
\end{split}
\end{equation} 
where $E^\text{R}_{mn}$ is the unknown amplitude of the $m$th reflected spatial-temporal diffracted orders in region 1, with the wavevectors $k_{x,mn}$ and $k_{z,mn}$. The total electric field in region 3 reads
\begin{equation}\label{eqa:ET}
\begin{split}
\mathbf{E}_\text{3} = \mathbf{\hat{y}} \sum_{m,n} E^\text{T}_{mn} e^{i ( k''_{x,mn} x +k''_{z,mn} z -\omega_{n} t)},
\end{split}
\end{equation} 
where $E^\text{T}_{mn}$ is the amplitude of the $m$th transmitted spatial-temporal diffracted order in region 3, with the wavevectors $k''_{x,mn}$ and $k''_{z,mn}$. To determine the unknown field coefficients of the backward and forward diffracted orders, $E^\text{R}_{mn}$ and $E^\text{T}_{mn}$, we enforce the continuity of the tangential electric and magnetic fields at the boundaries of the grating at $z=0$ and $z=d$. The electric field continuity condition between regions 1 and 2 at $z=0$, ${E_{1y}}(x,0,t) = {E_{2y}}(x,0,t)$, using~\eqref{eqa:Em} and~\eqref{eqa:ER}, reduces to
\begin{equation}\label{eqa:EBC_forw_12}
\delta_{n0} E_{0} e^{i  k_{x,\text{i}} x }+ \sum_{m,n} E^\text{R}_{mn} e^{i  k_{x,mn} x }=
 \sum_{p,m,n} E'_{pmn} e^{i k'_{x,pmn} x },
\end{equation}
%
%

\noindent and the corresponding magnetic field continuity condition, i.e., ${H_{1x}}(x,0,t) = {H_{2x}}(x,0,t)$, may be formulated as
\begin{equation}\label{eqa:HBC_forw_12}
\begin{split}
&\cos(\theta_\text{i})   \delta_{m0} \delta_{n0} E_0 e^{i  k_{x,\text{i}} x } - \cos(\theta_{mn}) E_{mn}^\text{R} e^{i  k_{x,mn} x }\\
& =\eta_1  \sum_{p } \frac{k'_{z,pmn}}{k'_{pmn}} \frac{E'_{pmn}}{\eta' } e^{i k'_{x,pmn} x },
\end{split}
\end{equation}

The electric field continuity condition between regions 2 and 3 at $z=d$, ${E_{2y}}(x,d,t) = {E_{3y}}(x,d,t)$, reduces to
\begin{equation}
E^\text{T}_{mn}=
\sum_{p} E'_{pmn} e^{i \left( [k'_{x,pmn}-k''_{x,mn}] x +[k'_{z,pmn}-k''_{z,mn}] d \right)} ,
\label{eqa:EBC_forw_23}
\end{equation}
while the corresponding tangential magnetic field continuity condition between regions 2 and 3 at $z=d$, ${H_{2x}}(x,d,t) = {H_{3x}}(x,d,t)$, reads
\begin{equation}\label{eqa:HBC_forw_23}
\begin{split}
 \sum_{p } \frac{E'_{pmn}}{\eta' } &e^{i \left( [k'_{x,pmn}- k''_{x,mn}] x + [k'_{z,pmn}-k''_{z,mn}] d \right)}\\
 &  = \frac{\cos(\theta_{mn})}{\cos(\theta_\text{i})} \frac{E_{mn}^\text{T}}{\eta'' } 
 \end{split} 
\end{equation}

\begin{figure}
	\begin{center}
		\includegraphics[width=1\columnwidth]{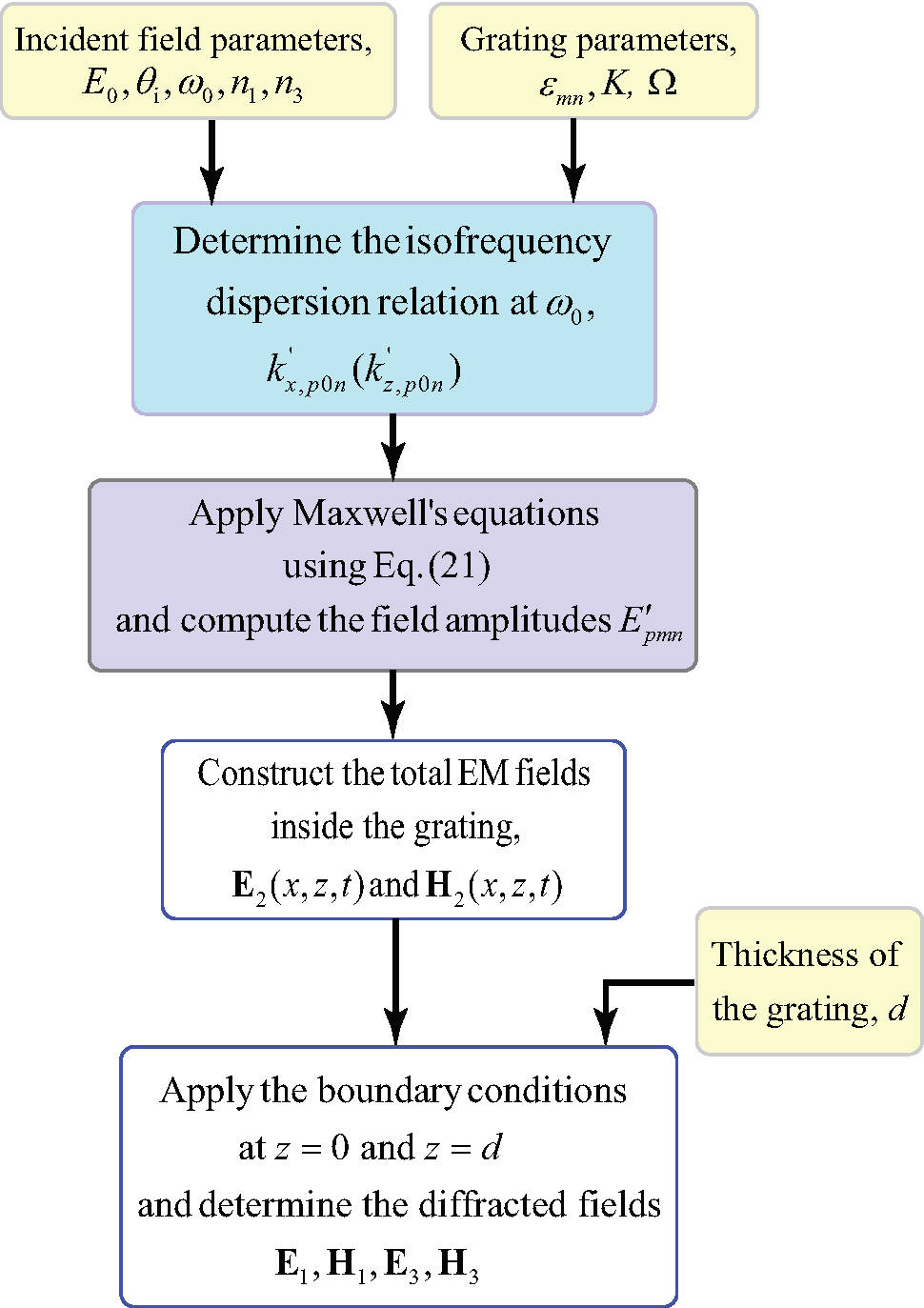} 
		\caption{Procedure for deriving the scattered electromagnetic fields inside and outside of a STP grating.} 
		\label{Fig:filed_sol}
	\end{center}
\end{figure}

Solving the above four equations, i.e., Eqs.~\eqref{eqa:EBC_forw_12}-~\eqref{eqa:HBC_forw_23}, together provides the four unknown field amplitudes, i.e., the forward and backward field amplitudes inside the grating ($E'^+_{pmn}$ and $E'^-_{pmn}$), and the reflected and transmitted field amplitudes outside the grating ($E_{mn}^\text{R}$ and $E_{mn}^\text{T}$). Figure~\ref{Fig:filed_sol} overviews the procedure for determining unknown field amplitudes and the dispersion relation of a STP grating.

\section{Illustrative Examples}\label{sec:num}
\subsection{Conventional Spatially Periodic Static Diffraction Grating}\label{sec:conv}

\begin{figure}
	\includegraphics[width=0.99\columnwidth]{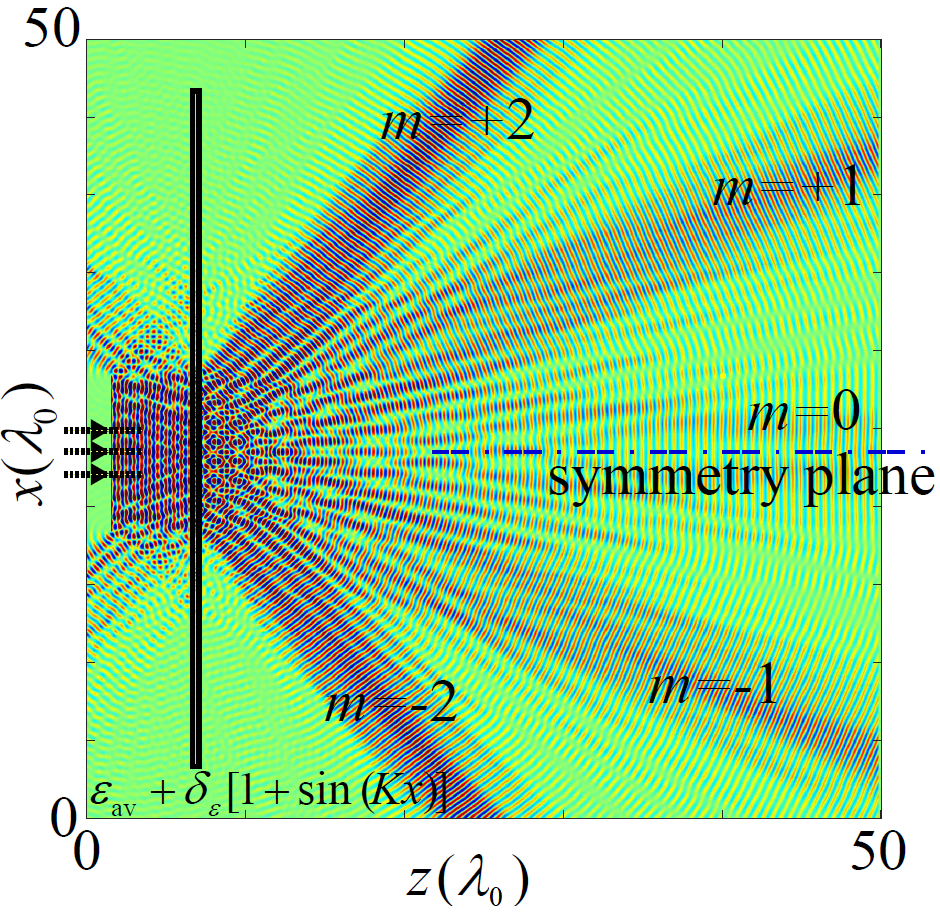}
	\caption{FDTD numerical simulation results of the $y$-component of the electric field for the diffraction from a conventional spatially periodic time-invariant grating ($\Omega=0$), with $\theta_\text{i}=0^\circ$, $\omega_0=2\pi\times10$~GHz, $\delta_{\epsilon}=0.5$, $K=0.4 k_0$, $d=0.8 \lambda_0$.}
	\label{fig:time_conv}
\end{figure}

For the sake of comparison, we first investigate the diffraction from a conventional planar spatially periodic (static) diffraction gratings~\cite{tamir1964wave,burckhardt1966diffraction,gaylord1982planar,gaylord1985analysis}. Such a static grating assumes a sinusoidal relative electric permittivity in the region from $z = 0$ to $z = d$ given by
\begin{equation}\label{eqa:eps1}
n_\text{gr}^2(x)= \epsilon_\text{av} +\delta_\epsilon  [1+\sin(Kx)],
\end{equation} 
and interfaced with two semi-infinite dielectric regions, characterized with refractive indices $n_1$ and $n_3$, respectively. In Eq.~\eqref{eqa:eps1}, $\delta_{\epsilon}$ represents the modulation strength. Figure~\ref{fig:time_conv} shows the time domain FDTD simulation results for the diffraction from a conventional spatially periodic grating with $\theta_\text{i}=0^\circ$, $\omega_0=2\pi\times10$~GHz, $\delta_{\epsilon}=0.5$, $\Omega=0$, $K=0.4 k_0$, $d=0.8 \lambda_0$. It may be seen from this figure that, for a monochromatic incident wave, all spatial diffracted orders possess the same wavelength (frequency). Another observed phenomenon is that, since the grating is "undirectional", the diffraction pattern for a normal incidence ($\theta_{\text{i}}=0$) is symmetric with respect to the $x$ axis. Table~\ref{tab:table0} lists the analytical results, derived from~\eqref{eqa:refl_trans_angl}, for the diffraction angles $\theta_{m}$ (in degrees) of the conventional transmissive space periodic diffraction grating in Fig.~\ref{Fig:grating_conv}.

\begin{table}
	\centering
	\caption{Analytical results for diffraction angles $\theta_{m}$ (in degrees) of the transmissive conventional spatially periodic time-invariant grating ($\Omega=0$), corresponding to the FDTD numerical simulation results in Fig.~\ref{fig:time_conv}.} 
	\label{tab:table0}
	\begin{tabular}{|c|c|c|c|c|c|c|  }
		\hline
		 \multicolumn{7}{c}{$m$} \\ 
			$-3$& $-2$& $-1$& $0$& $+1$ &$+2$&$+3$\\
		\hline 	\hline
		Ev. &  -53.1 & -23.58    &    0  & 23.58 & 53.1 & Ev.\\ \hline
	\end{tabular}
\end{table}	

\subsection{Asymmetric Pattern of a STP Diffraction Grating}\label{sec:asym_patt}
Next, we demonstrate the diffraction from a planar STP (dynamic) diffraction grating. As a particular case, which is practical and common, we study the grating with a sinusoidal relative electric permittivity in the region from $z = 0$ to $z = d$ (see animation in Supplemental Material~\cite{supplementalMaterial_2019}) given by
\begin{equation}\label{eqa:eps2}
n_\text{gr}^2(x,t)= \epsilon_\text{av} +\delta_\epsilon  [1+\sin(Kx-\Omega t)].
\end{equation} 

To compute the solution derived in Sec.~\ref{eqa:fields}, we shall write the expression in~\eqref{eqa:eps2} in terms of its spatial-temporal Fourier components, considering the general form given in Eq.~\eqref{eqa:index}, i.e.,
\begin{subequations}\label{eqa:eps_sin_exp}
	\begin{equation}
n_\text{gr}^2(x,t) = {\epsilon_{-1,-1}}{e^{ -j(Kx-\Omega t )}} + \epsilon_{00} + \epsilon_{11}e^{ + j(Kx-\Omega t )},
\end{equation}
with
\begin{equation}\label{eqa:eps3}
	\epsilon_{11}=-\epsilon_{-1,-1}= \delta_\epsilon/2i\quad\text{and}\quad \epsilon_{00}=\epsilon_\text{av}+\delta_\epsilon.
\end{equation}
\end{subequations}

We next insert the nonzero terms of the permittivity, given in~\eqref{eqa:eps3}, into Eq.~\eqref{eqa:Max2_c}, and determine the electromagnetic fields inside the grating, dispersion relation~\cite{Taravati_PRB_2017,Taravati_PRAp_2018,Taravati_Kishk_TAP_2019}, and the diffracted fields. Table~\ref{tab:table1} lists the analytical results for the diffraction angles $\theta_{mn}$ (in degrees) of the transmissive STP diffraction grating, for normal incidence of a monochromatic wave, $\theta_\text{i}=0^\circ$, $\omega_\text{i}=\omega_0=2\pi\times10$~GHz, where $\delta_{\epsilon}=0.5$, $\Omega=0.28\omega_0$, $K=0.4 k_0$, $d=0.8 \lambda_0$. 

\begin{figure}
	\subfigure[]{\label{Fig:aa}
		\includegraphics[width=0.99\columnwidth]{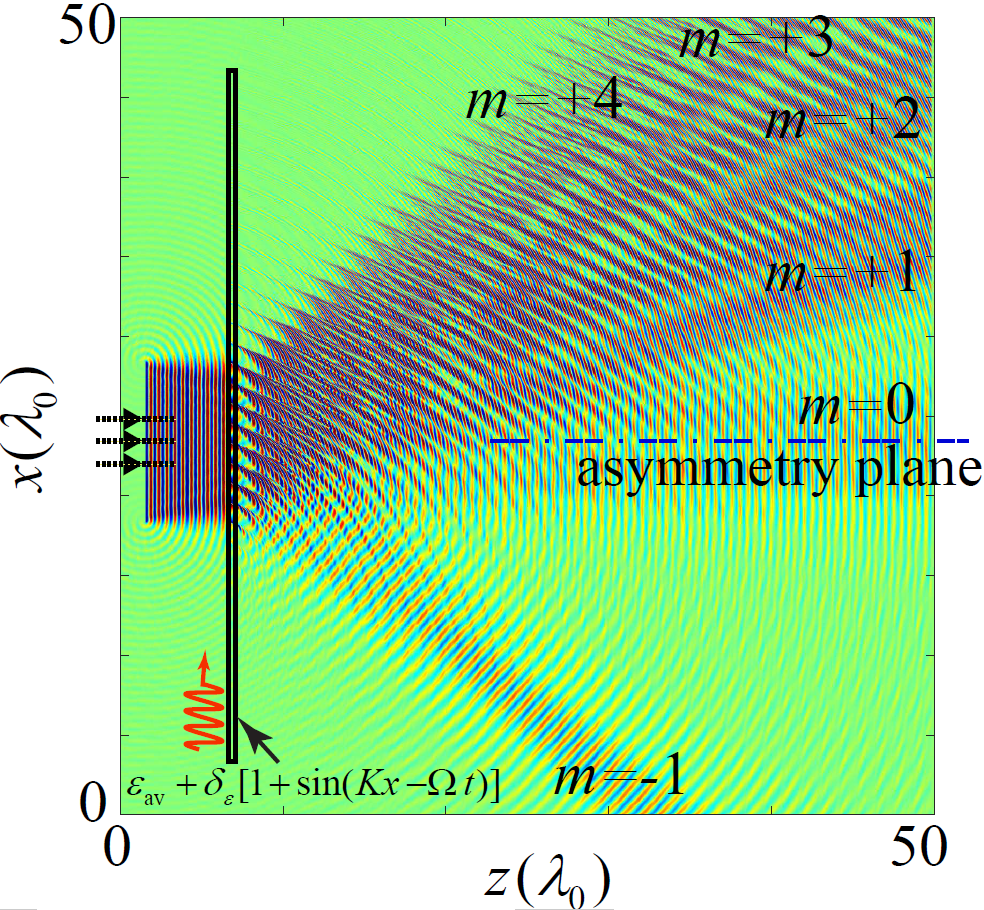}} 
	\subfigure[]{\label{Fig:bb}
		\includegraphics[width=0.48\columnwidth]{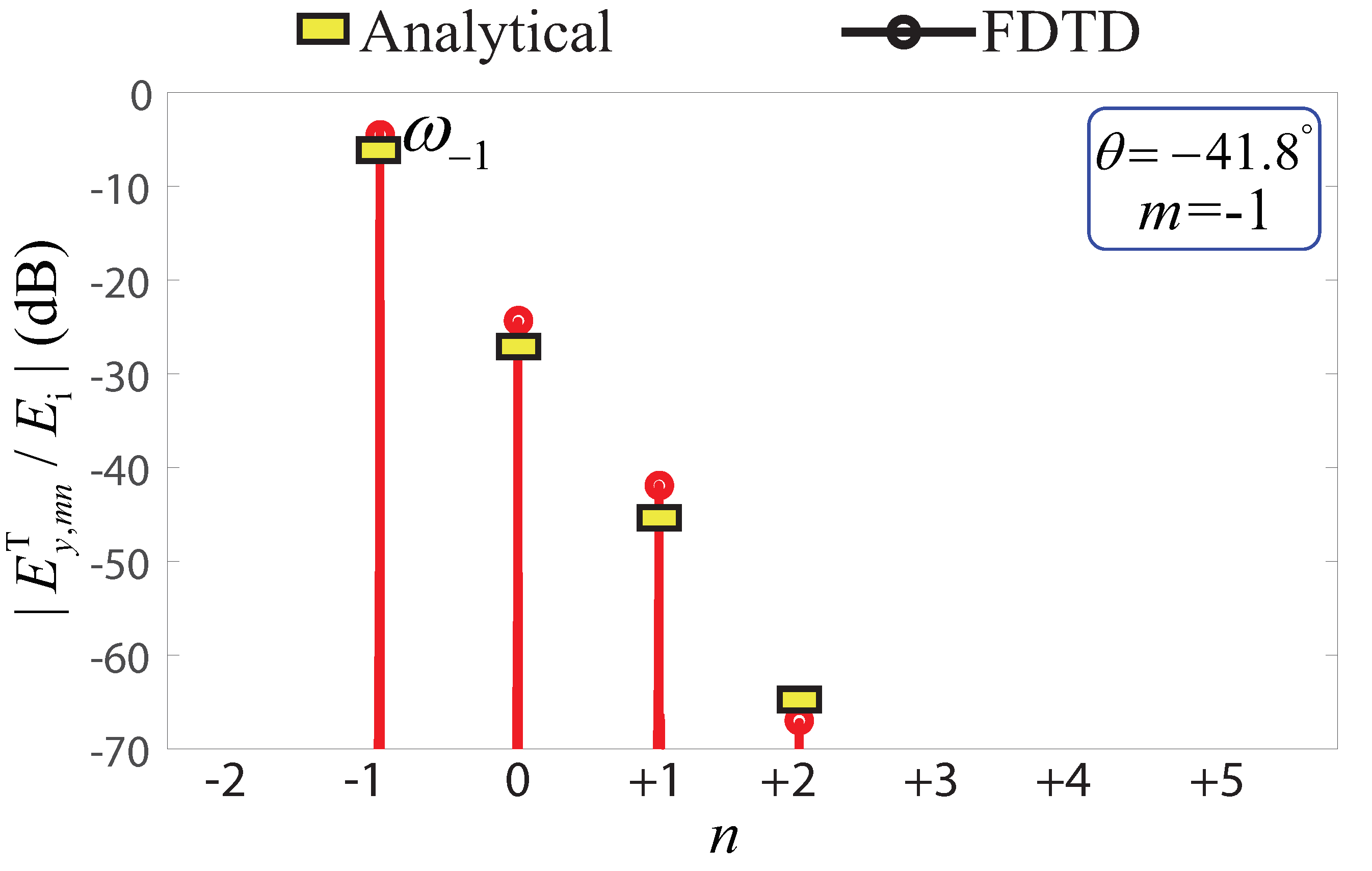}}
	\subfigure[]{\label{Fig:cc}
		\includegraphics[width=0.48\columnwidth]{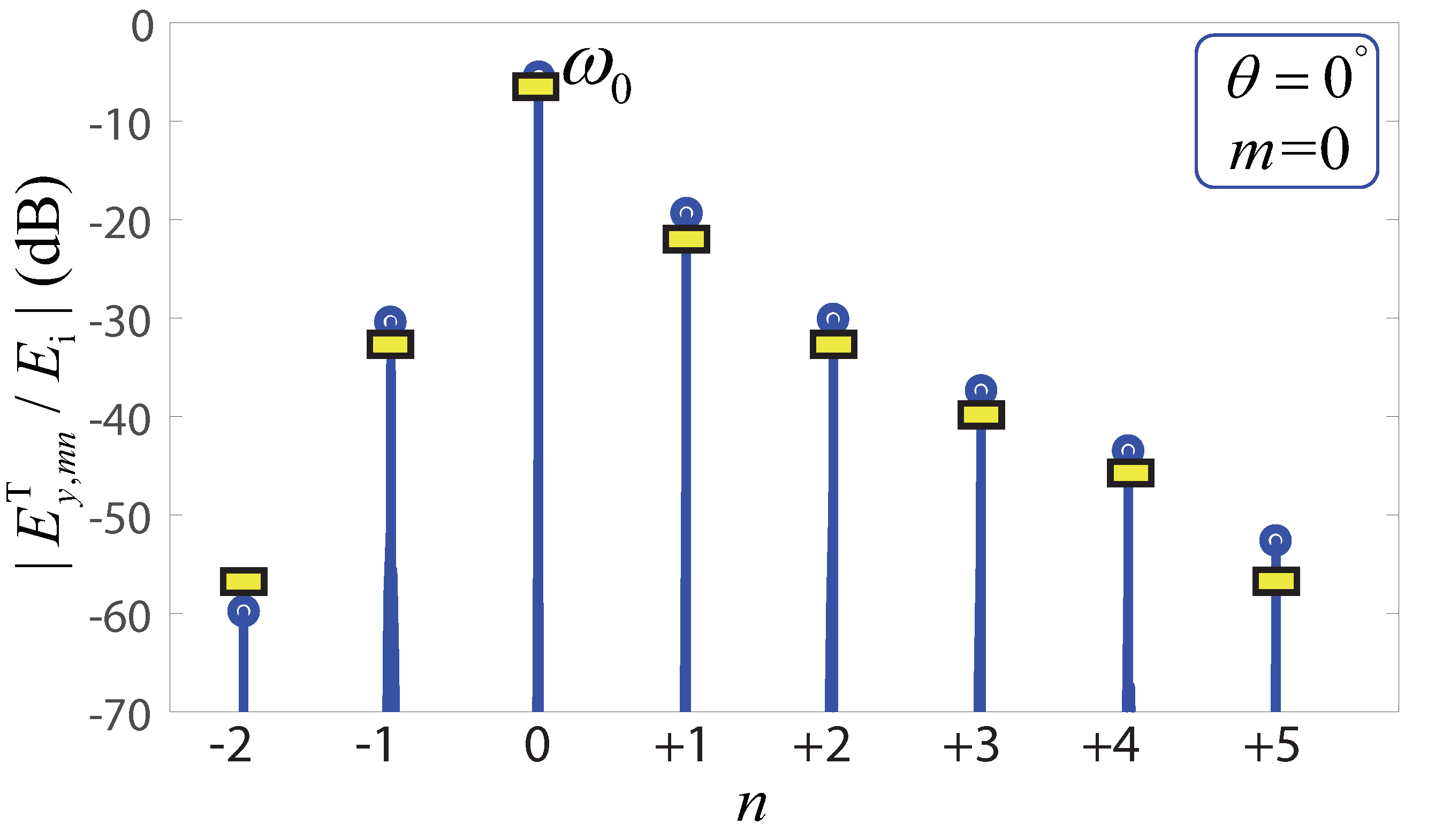}}
	\subfigure[]{\label{Fig:dd}
		\includegraphics[width=0.48\columnwidth]{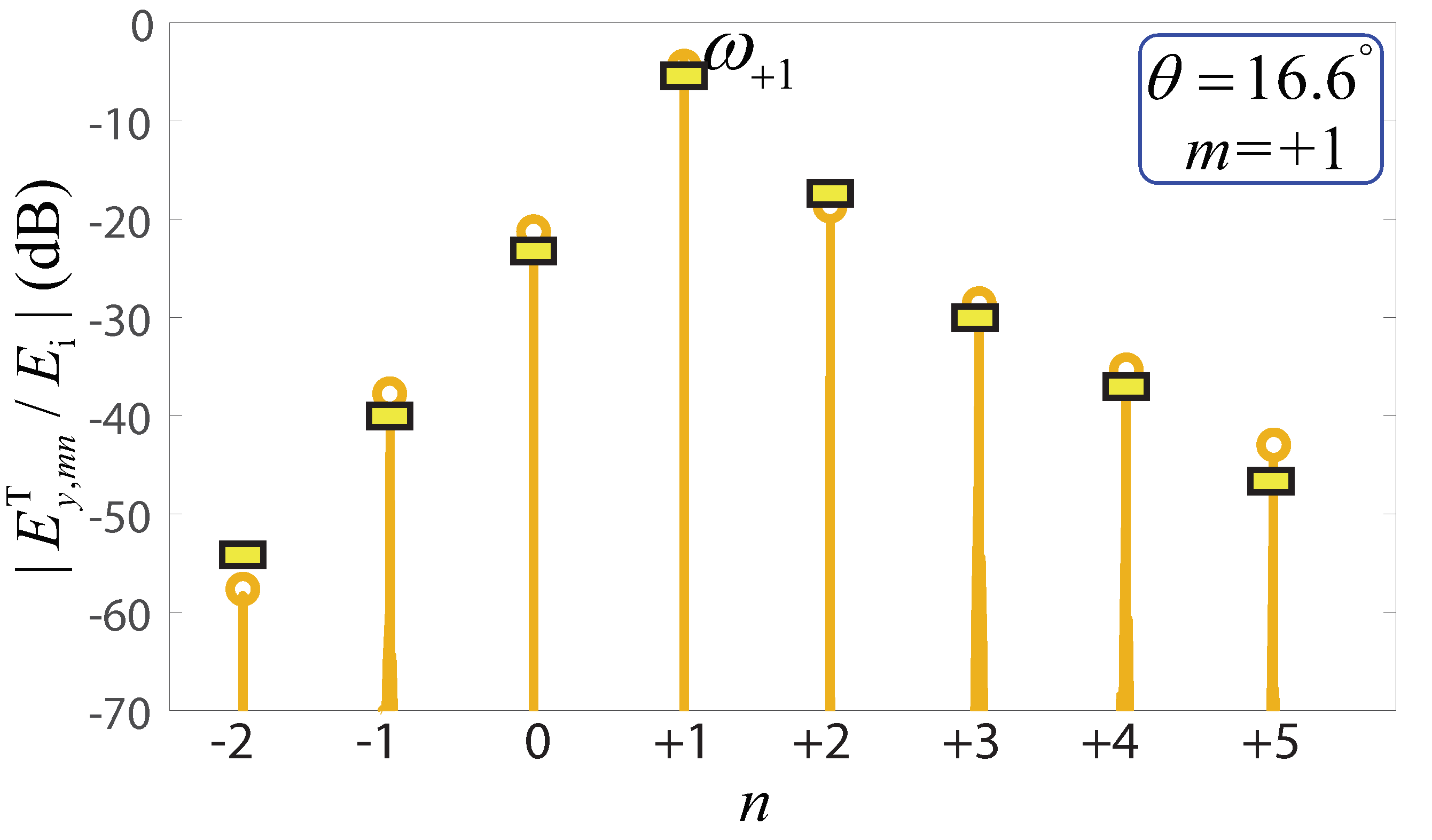}}
	\subfigure[]{\label{Fig:ee}
		\includegraphics[width=0.48\columnwidth]{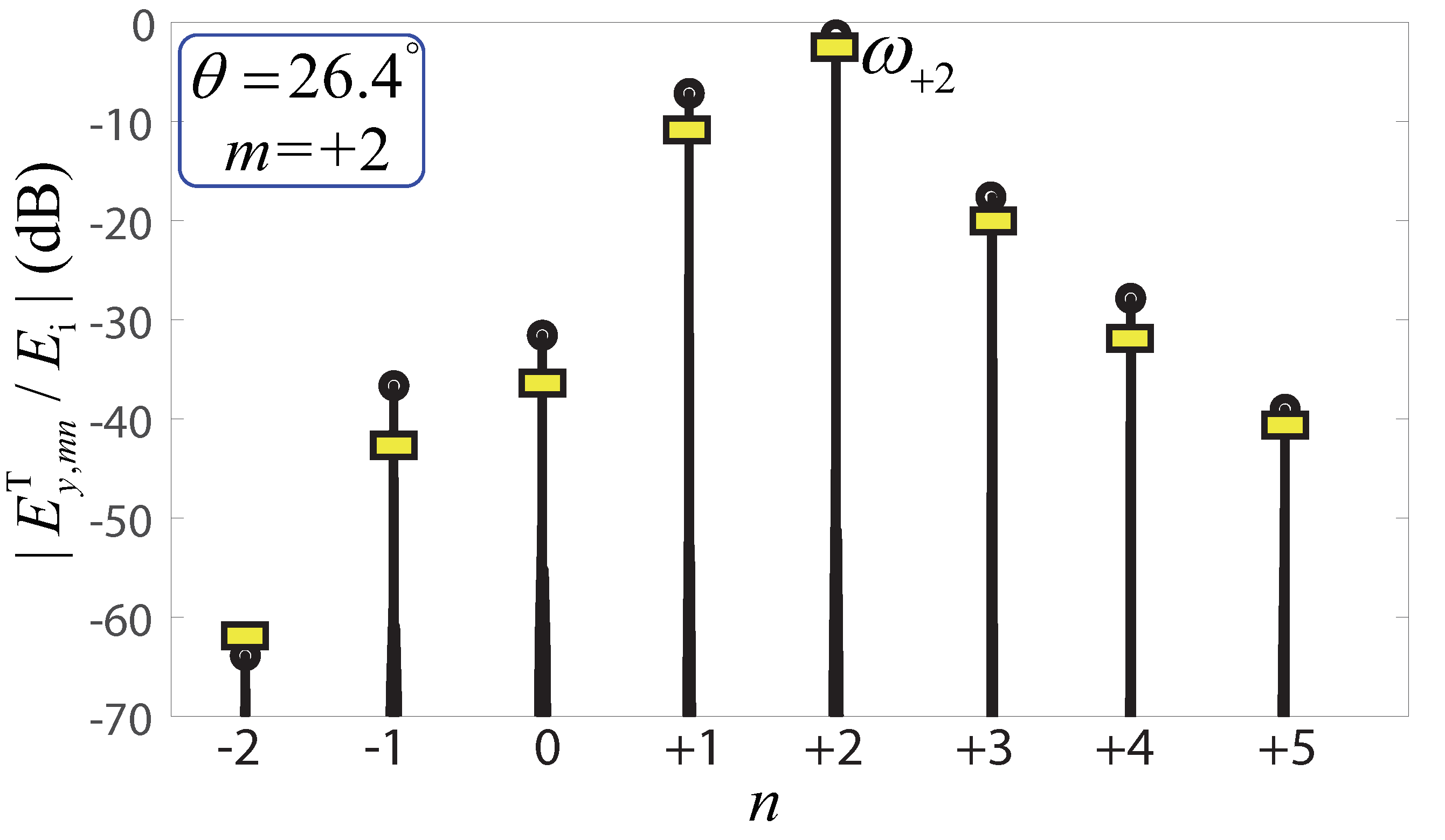}}
	\subfigure[]{\label{Fig:ff}
		\includegraphics[width=0.48\columnwidth]{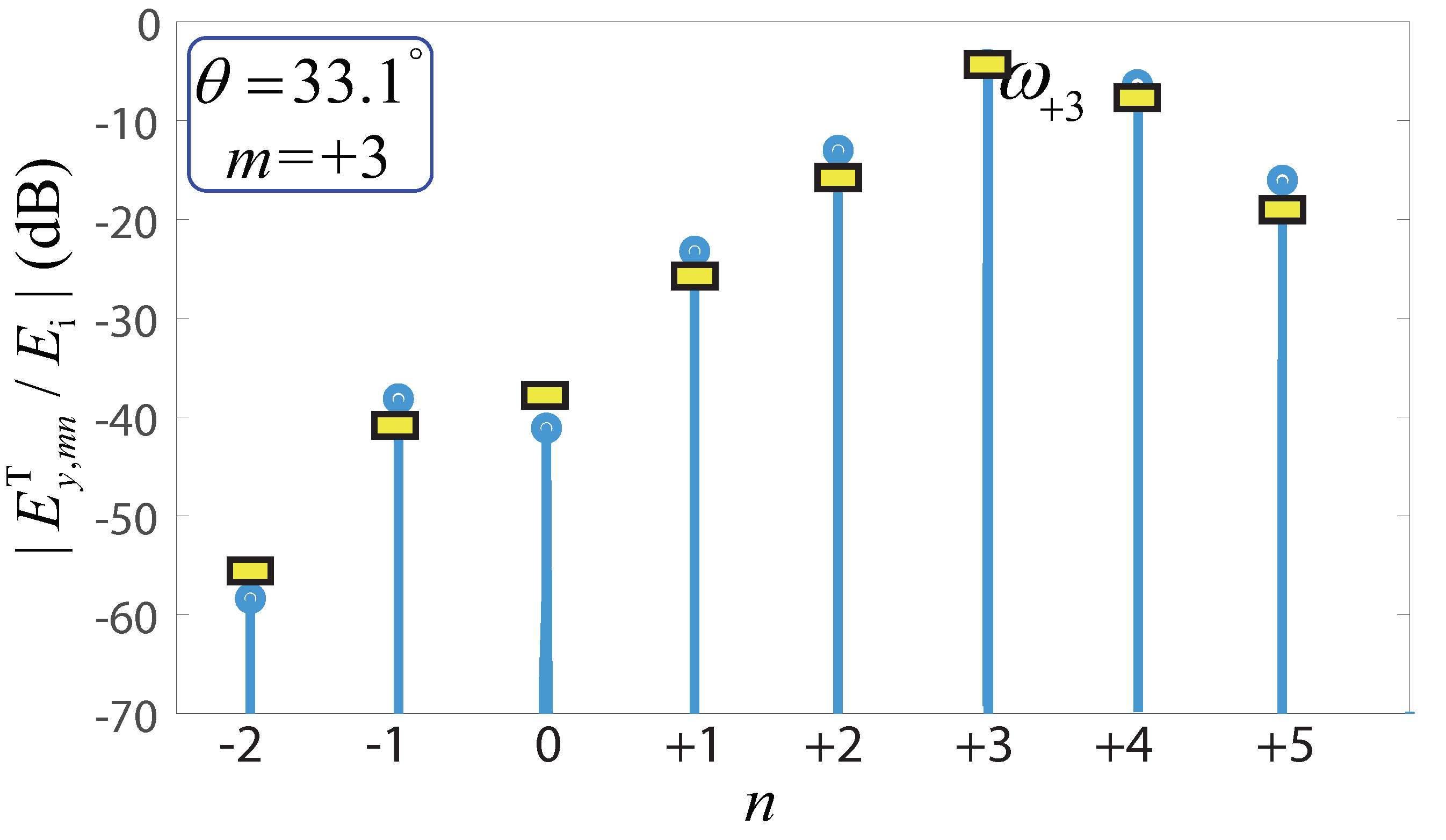}}
	\subfigure[]{\label{Fig:gg}
		\includegraphics[width=0.48\columnwidth]{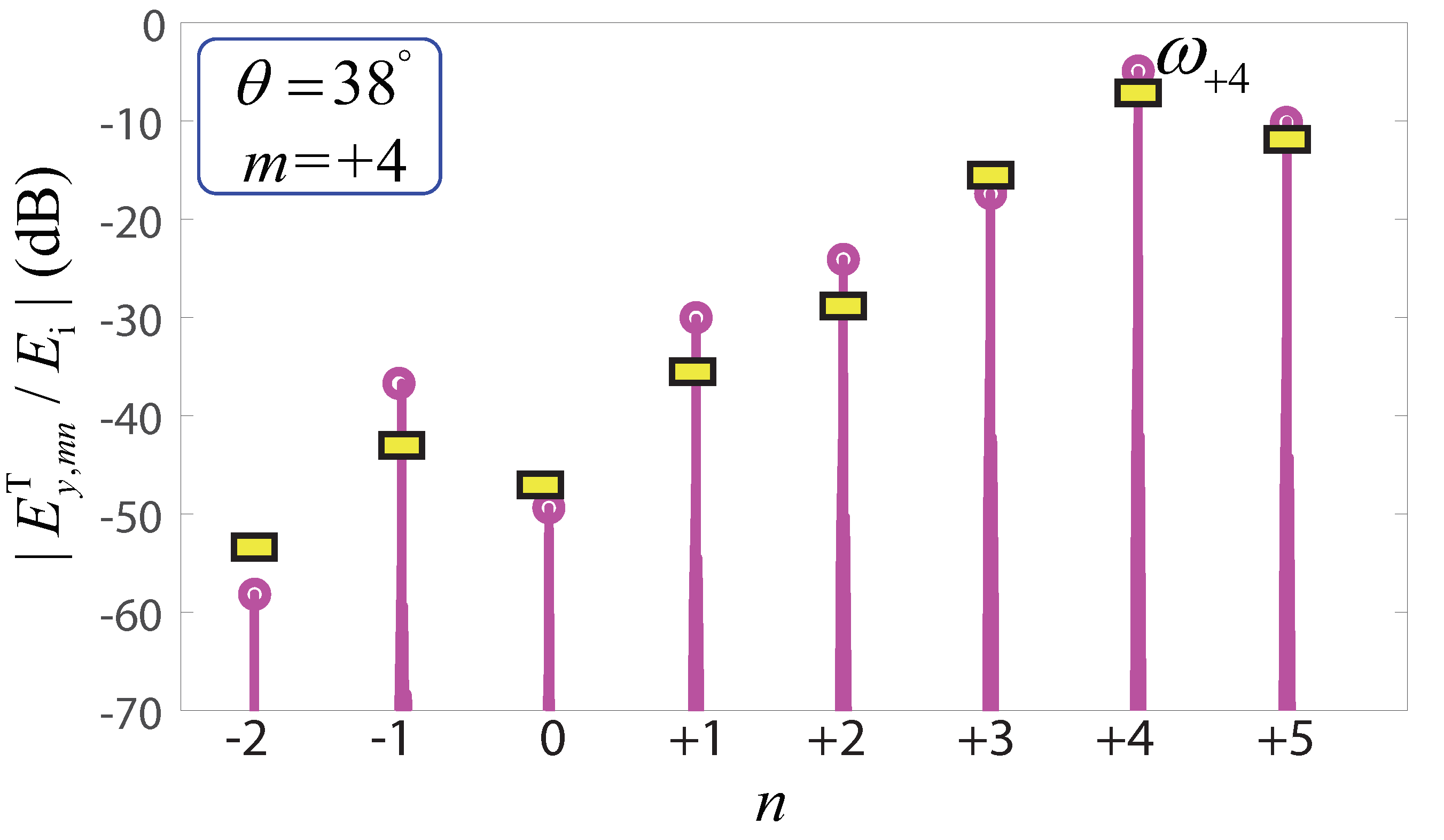}} 
	\caption{Analytical and FDTD simulation results for the spatial-temporal diffraction from a STP grating, for normal incidence of a plane wave ($\theta_\text{i}=0^\circ$) with temporal frequency $\omega_0=2\pi\times10$~GHz, where $\delta_{\epsilon}=0.5$, $\Omega=0.28\omega_0$, $K=0.4 k_0$, $d=0.8 \lambda_0$. (a) FDTD simulation time domain response of the $y$-component of the transmitted electric field, i.e., $E^\text{T}_{y,mn}$ (see animation in Supplemental Material~\cite{supplementalMaterial_2019}). (b)-(g) Frequency domain responses for (b) $m=-1$. (c) $m=0$. (d) $m=+1$. (e) $m=+2$. (f) $m=+3$. (g) $m=+4$. The analytical results for the angles of diffraction are listed in Table~\ref{tab:table1}.}
	\label{fig:Trans_FDTD}
\end{figure}

Figure~\ref{Fig:aa} shows the corresponding time domain FDTD simulation results for the diffraction from this STP grating. We observe from this figure that, in contrast with the conventional case in Fig.~\ref{fig:time_conv}, the diffraction pattern of the STP grating is asymmetric with respect to the $x$ axis. The second observed phenomenon, as expected, is that the diffracted orders possess different wavelengths, which correspond to different frequencies. From Fig.~\ref{Fig:aa}, one may conclude that each diffracted order is attributed to a single frequency. However, this is not true. To see the exact phenomenon, we shall look at the frequency spectrum of the diffracted orders, by performing a fast Fourier transform of the transmitted diffracted orders at different angles. Figures~\ref{Fig:bb} to~\ref{Fig:gg} plot the analytical and FDTD numerical simulation frequency domain responses for the $m=-1$ to $m=+4$ diffracted orders, corresponding to the analytical results listed in Tab.~\ref{tab:table1}. These figures show that each diffracted spatial order includes an infinite set of temporal harmonics, $\omega_n=\omega_0+n \Omega$, with $n$ being any integers.
\begin{table}
	\centering
	\caption{Analytical results for the diffraction angles $\theta_{mn}$ (in degrees) of the transmissive STP diffraction grating, where the FDTD numerical simulation results are given in Fig.~\ref{fig:Trans_FDTD}.} 
	\label{tab:table1}
	\begin{tabular}{|c||c|c|c|c|c|c|c|  }
		\hline
		& \multicolumn{7}{c}{$m$} \\ 
		&$-3$& $-2$& $-1$& $0$& $+1$ &$+2$&$+3$\\
		\hline 	\hline
		$n=-3$& Ev. &Ev.&Ev.&0&Ev. &Ev. &Ev.\\ \hline
		$n=-2$& Ev.&Ev.& -65.4 &0& 65.4 &Ev.&Ev.\\ \hline
		$n=-1$&Ev. & Ev. &\underline{\textbf{-33.7}}   &0 & 33.7 &Ev. & Ev.\\ \hline
		$n=0$&-70 &  -38.7 & -18.2    &    \underline{\textbf{0}}  & 18.2 & 38.7 & 70\\ \hline
		$n=+1$&  -59  &   -34.85 &   -16.6 &     0 & \underline{\textbf{16.6}}  &  34.85 &  59\\
		\hline
		$n=+2$& -50.3  &   -30.8 &   -14.86 &         0 & 14.86 &  \underline{\textbf{30.8}} &  50.3  \\\hline
		$n=+3$& -40.7.7 &   -25.8 &   -12.55 &         0 & 12.55 &  25.8 & \underline{\textbf{40.7}} \\\hline
	\end{tabular}
\end{table}	

\vspace{6mm}
\subsubsection{Effect of the grating thickness}
It is of great interest to investigate the effect of the thickness of the STP grating ($d$) on the generation of space and time diffraction orders and the grating efficiency. In general, diffraction gratings may be classified in two main categories, i.e., \textit{thin} and \textit{thick} gratings, each of which exhibiting its own angular and wavelength selectivity characteristics. The thin gratings usually result in Raman-Nath regime diffraction, where multiple diffracted orders are produced. In contrast, the thick gratings usually result in Bragg regime diffraction, where only one single diffracted order is produced. Following the procedure described in~\cite{hutley1982diffraction,gaylord1985analysis}, we characterize these two diffraction regimes, i.e., the Bragg and Raman-Nath regimes, by the dimensionless parameter 

\begin{equation}
Q_n=\frac{v_\text{r} K^2 d}{(\omega_0+n \Omega) \cos(\theta'_n) }
\end{equation}

The grating strength parameter is represented by 
\begin{equation}
\gamma_n=\frac{\delta_\epsilon}{\epsilon_\text{av}}\frac{ d (\omega_0+n \Omega)}{4 v_\text{r}  \cos(\theta'_n) }
\end{equation}
for TE polarization, and
\begin{equation}
\gamma_n=\frac{\delta_\epsilon}{\epsilon_\text{av}}\frac{ d (\omega_0+n \Omega) \cos(2\theta'_n)}{4 v_\text{r}  \cos(\theta'_n) }
\end{equation}
for TM polarization.

\begin{figure*}
		\subfigure[]{\label{Fig:Raman}
		\includegraphics[width=1\columnwidth]{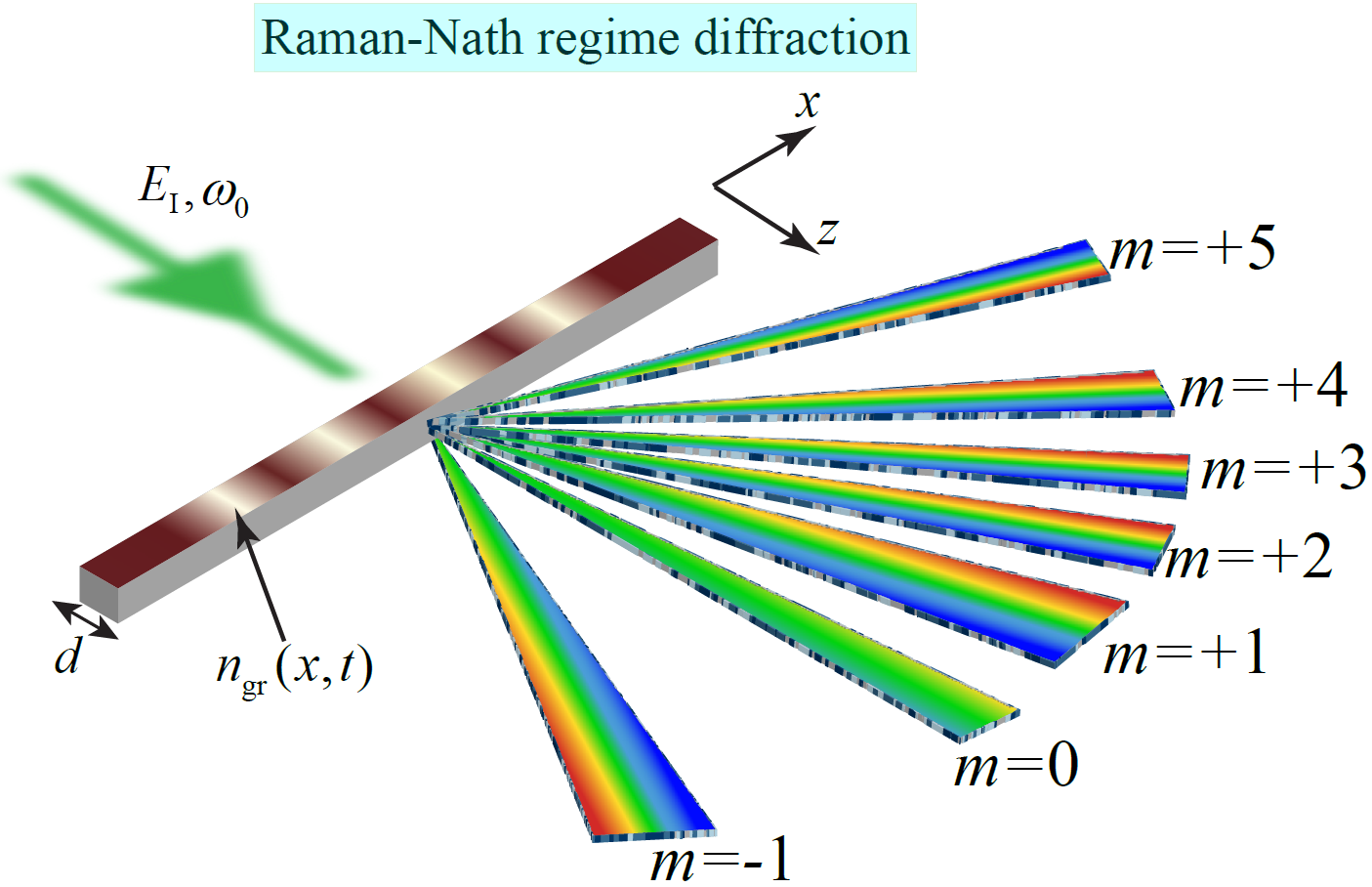}}
	\subfigure[]{\label{Fig:Raman_res}
		\includegraphics[width=1\columnwidth]{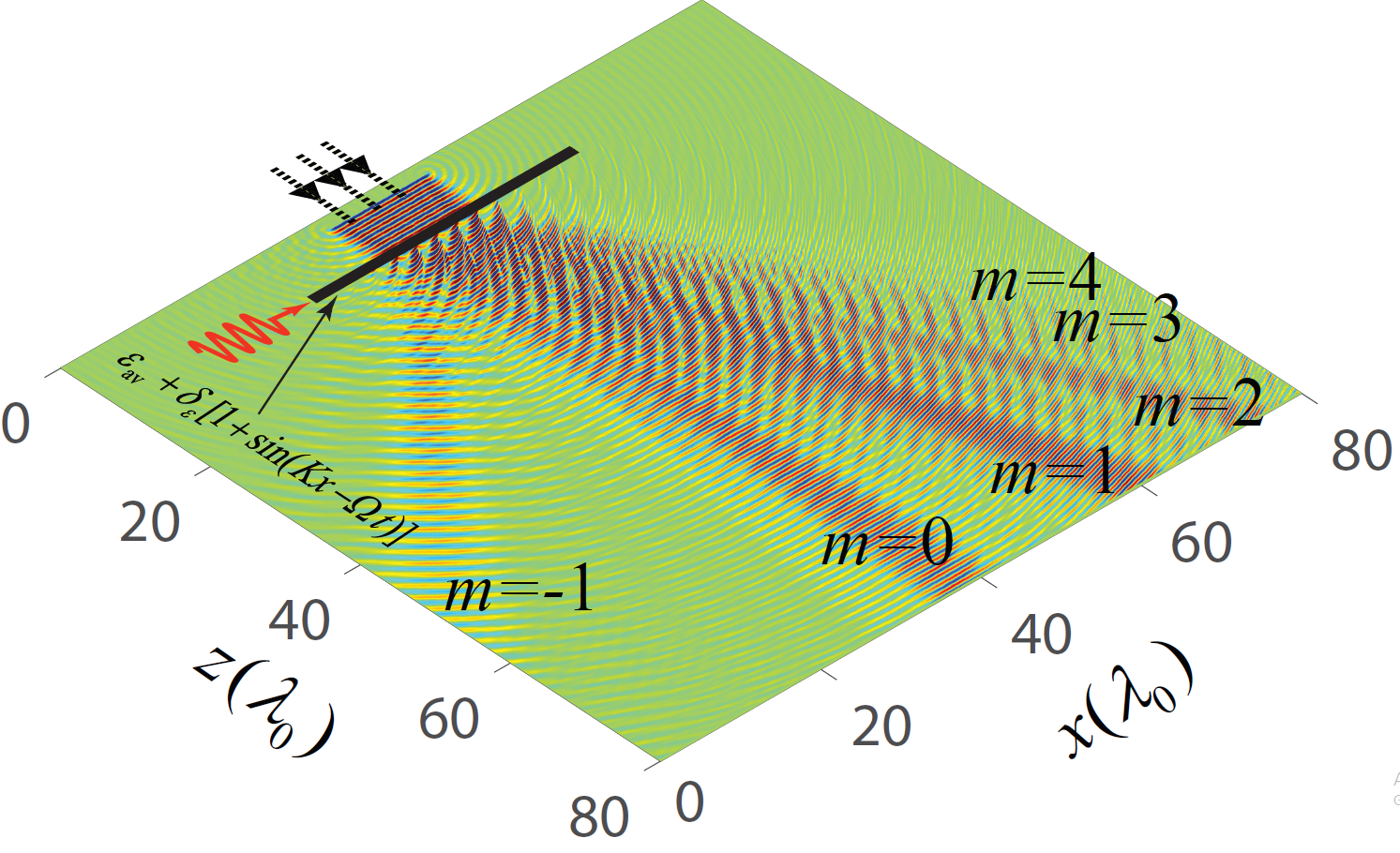}}
	\subfigure[]{\label{Fig:Bragg}
		\includegraphics[width=1\columnwidth]{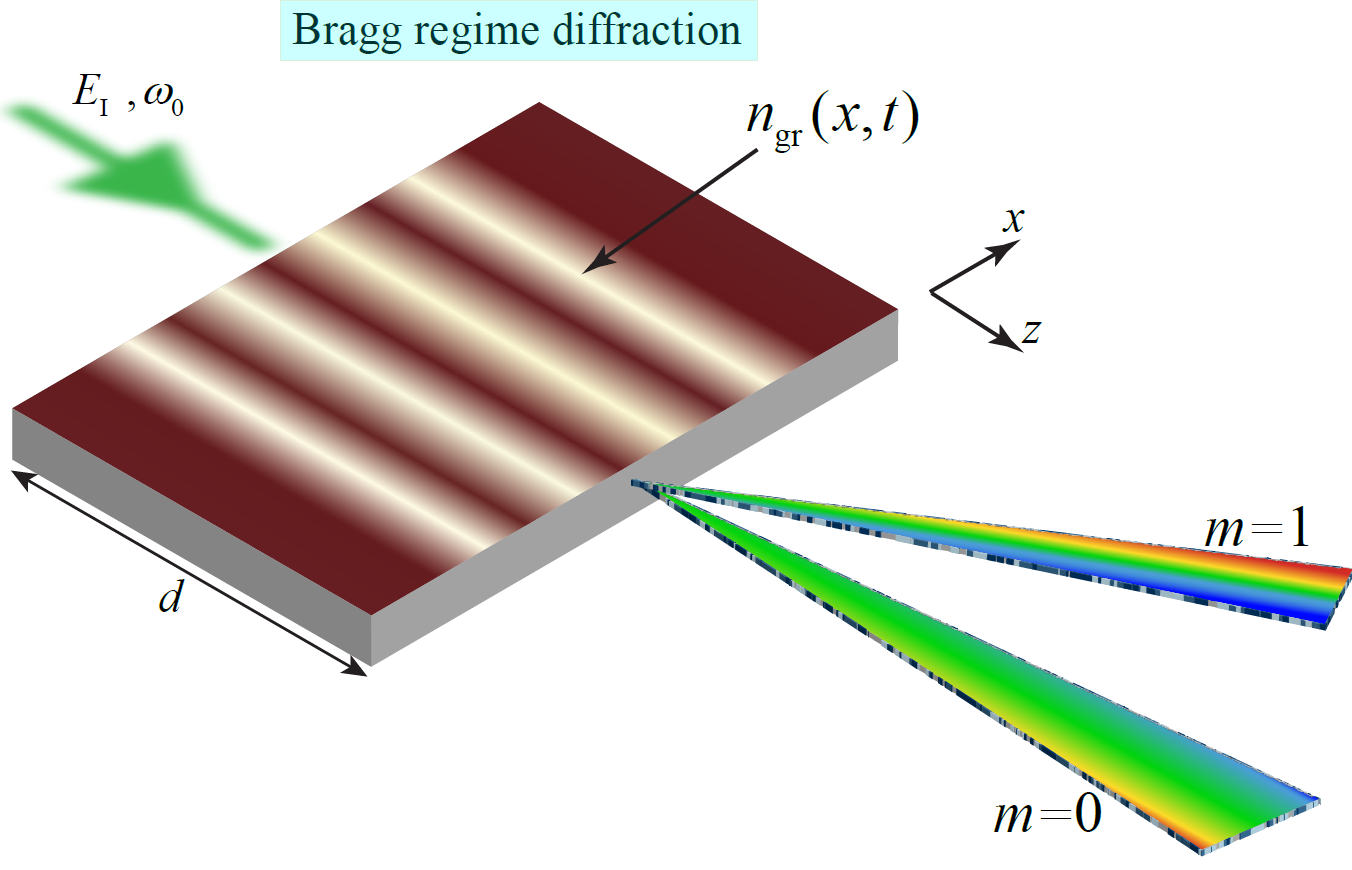}}
		\subfigure[]{\label{Fig:Bragg_res}
		\includegraphics[width=1\columnwidth]{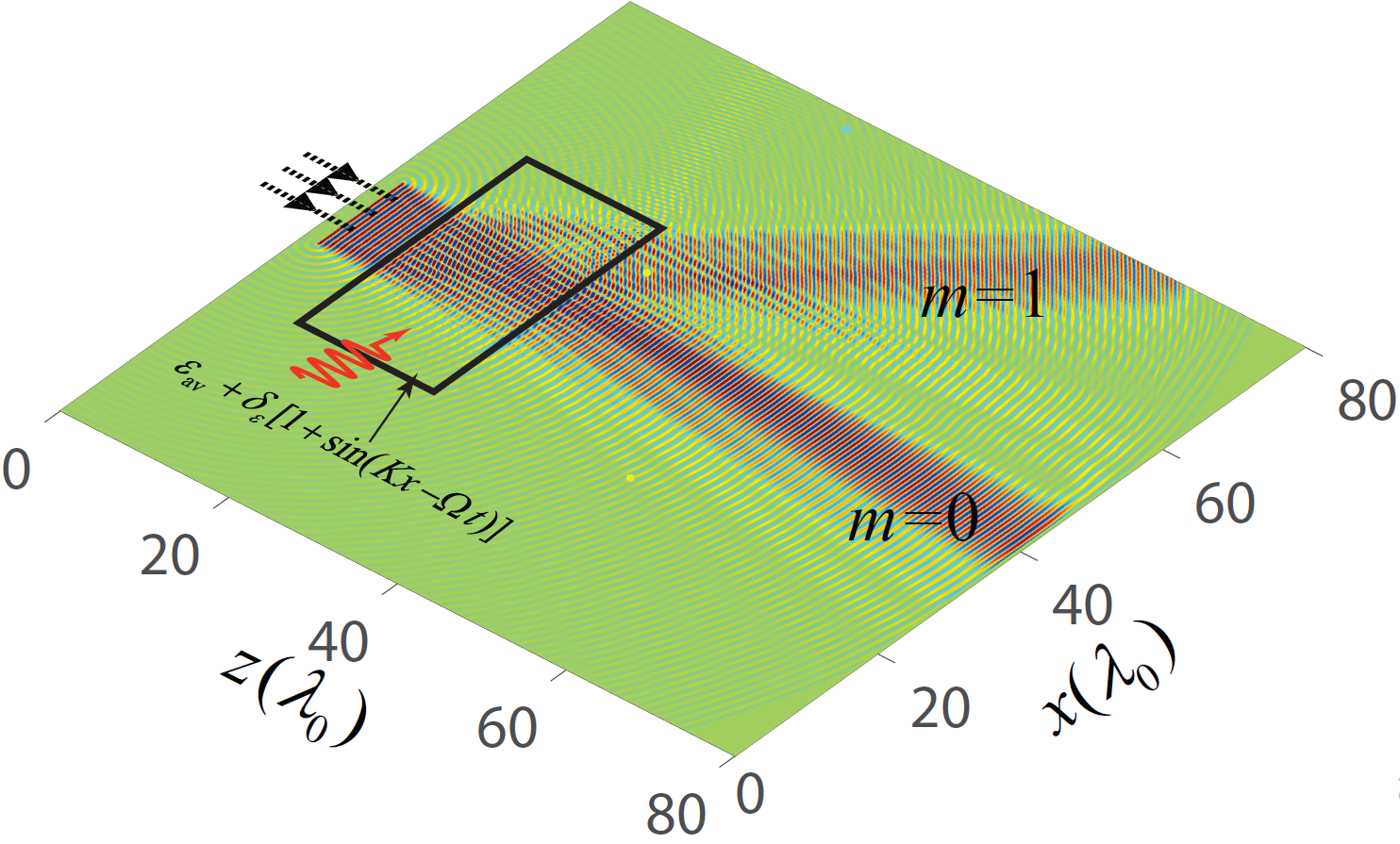}}
	\caption{Operation regimes of STP transmissive diffraction gratings, for normal incidence ($\theta_\text{i}=0^\circ$), $\omega_0=2\pi\times10$~GHz. (a) Raman-Nath regime diffraction of a thin grating, where $\Omega=0.4\omega_0$ and $K=0.4 k_0$, $\delta_{\epsilon}=0.5$ and $d=0.5 \lambda$. (b) Bragg regime diffraction of a thick grating, where $\Omega=0.347\omega_0$ and $K=0.867 k_0$, $\delta_{\epsilon}=0.1$ and $d=16 \lambda$.}
	\label{fig:Regimes}
\end{figure*}

\textit{Thin STP grating (Raman–Nath regime)}: The required condition for thin STP gratings exhibiting Raman-Nath regime
diffraction is represented by 
\begin{equation}
Q_n \gamma_n   \leq 1
\end{equation}

Thin gratings may be also characterized as gratings showing small angular and wavelength selectivity. As the incident wave is dephased (either in angle of incidence or in wavelength) from the Bragg condition, the diffraction efficiency decreases. The angular range or wavelength range for which the diffraction efficiency decreases to half of its
on-Bragg-angle value is determined by the thickness of the
grating $d$ expressed as a number of grating periods $\Lambda=2\pi/K$. For a
thin grating this number is reasonably chosen to be

\begin{equation}
Kd   \leq 20\pi
\end{equation}

Figure~\ref{Fig:Raman} shows a generic representation of the Raman-Nath regime diffraction in STP transmissive diffraction gratings, for normal incidence ($\theta_\text{i}=0^\circ$), $\omega_0=2\pi\times10$~GHz, $\Omega=0.4\omega_0$ and $K=0.4 k_0$, $\delta_{\epsilon}=0.5$ and $d=0.5 \lambda$. Figure~\ref{Fig:Raman_res} shows the numerical simulation results for Raman-Nath regime diffraction of the STP grating in Fig.~\ref{Fig:Raman}. Following the procedure described in~\cite{hutley1982diffraction,gaylord1985analysis} (for conventional spatially periodic gratings), for a thin transmissive STP grating operating in the Raman-Nath regime, the diffraction efficiency reads
\begin{equation}
\eta_{mn}=\frac{P_{mn}}{P_\text{inc}}=J_m^2(2\gamma_n)
\end{equation}
where $P_{mn}$ and $P_\text{inc}$ are the diffracted and incident powers, respectively, and where $J$ represents the, integer-order, ordinary Bessel function of the first kind.

\textit{Thick STP grating (Bragg regime)}: The Bragg regime diffraction may be achieved in thick gratings, where
then the required condition is 

\begin{equation}
\frac{Q_n}{2 \gamma_n}    \geq 10
\end{equation}

Thick gratings are capable of exhibiting strong angular and wavelength selectivity. A relatively small change in the angle of incidence from the Bragg angle or a relatively small change in the wavelength at the Bragg angle may result in a relatively strong dephasing, which in turn, decreases the diffraction efficiency. Thick grating behavior occurs when

\begin{equation}
Kd   \geq 20\pi
\end{equation}

Figure~\ref{Fig:Bragg} shows a generic representation of the Bragg regime diffraction in STP transmissive diffraction gratings, for normal incidence ($\theta_\text{i}=0^\circ$), $\omega_0=2\pi\times10$~GHz, $\Omega=0.4\omega_0$ and $K=0.4 k_0$, $\delta_{\epsilon}=0.5$ and $d=0.5 \lambda$. Figure~\ref{Fig:Bragg_res} shows the numerical simulation results for Bragg regime diffraction of the STP grating in Fig.~\ref{Fig:Bragg}. Following the procedure described in~\cite{hutley1982diffraction,gaylord1985analysis} (for conventional spatially periodic gratings), for a thick transmissive STP grating operating in the Bragg regime, the diffraction efficiency reads
\begin{equation}
\eta_{1n}=\sin^2(2\gamma_n)
\end{equation}

\subsection{Asymmetric and Nonreciprocal Response of Diffraction Orders in STP Gratings}\label{sec:nonr_asymm}
Over the past few years, there has been a surge of interest in the nonreciprocal~\cite{Fan_NPH_2009,Alu_PRB_2015,Fan_APL_2016,Fan_mats_2017,Taravati_PRB_2017,sounas2017non,Taravati_PRAp_2018,caloz2018electromagnetic,merkel2018dynamic} and asymmetric~\cite{fedotov2006asymmetric,singh2009terahertz,feng2011nonreciprocal,fan2012comment,wang2018extreme} wave transmission, reflection and absorption. Here, we investigate the realization of nonreciprocal, asymmetric and angle-asymmetric wave diffraction by STP gratings. The spatial diffraction of electromagnetic waves by natural media is reciprocal under reversal of the incident wave direction, whereas asymmetric and nonreciprocal spatial diffraction of electromagnetic waves has been recently achieved using different techniques~\cite{glass1990nonreciprocal,udalov2012nonreciprocal,guo2014nonreciprocal}. It should be noted that asymmetry and nonreciprocity in electromagnetic systems are different . The main difference is that asymmetric structures, which are linear and time-invariant, are constrained by the Lorentz reciprocity and cannot create optical isolators. For the sake of clarification, all time-invariant
linear systems, represented by symmetric electric
permittivity tensors and symmetric magnetic permeability tensors, are restricted by the Lorentz reciprocity theorem. Such systems are reciprocal as their scattering matrices are symmetric even if the electric
permittivity tensor or the magnetic permeability tensor are complex (system introduces gain or loss).  

\begin{figure}
		\subfigure[]{\label{Fig:asym}
			\includegraphics[width=1\columnwidth]{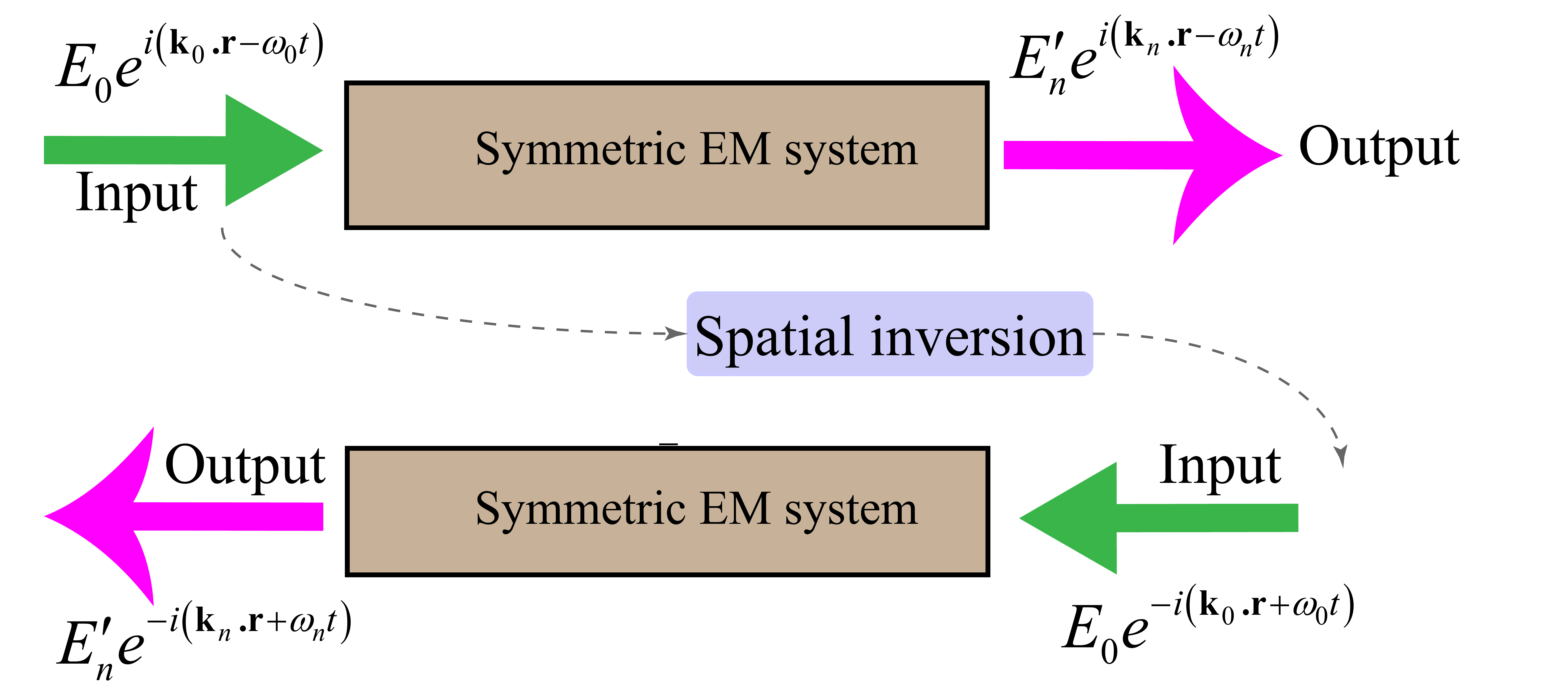}}
		\subfigure[]{\label{Fig:nonrec}
			\includegraphics[width=1\columnwidth]{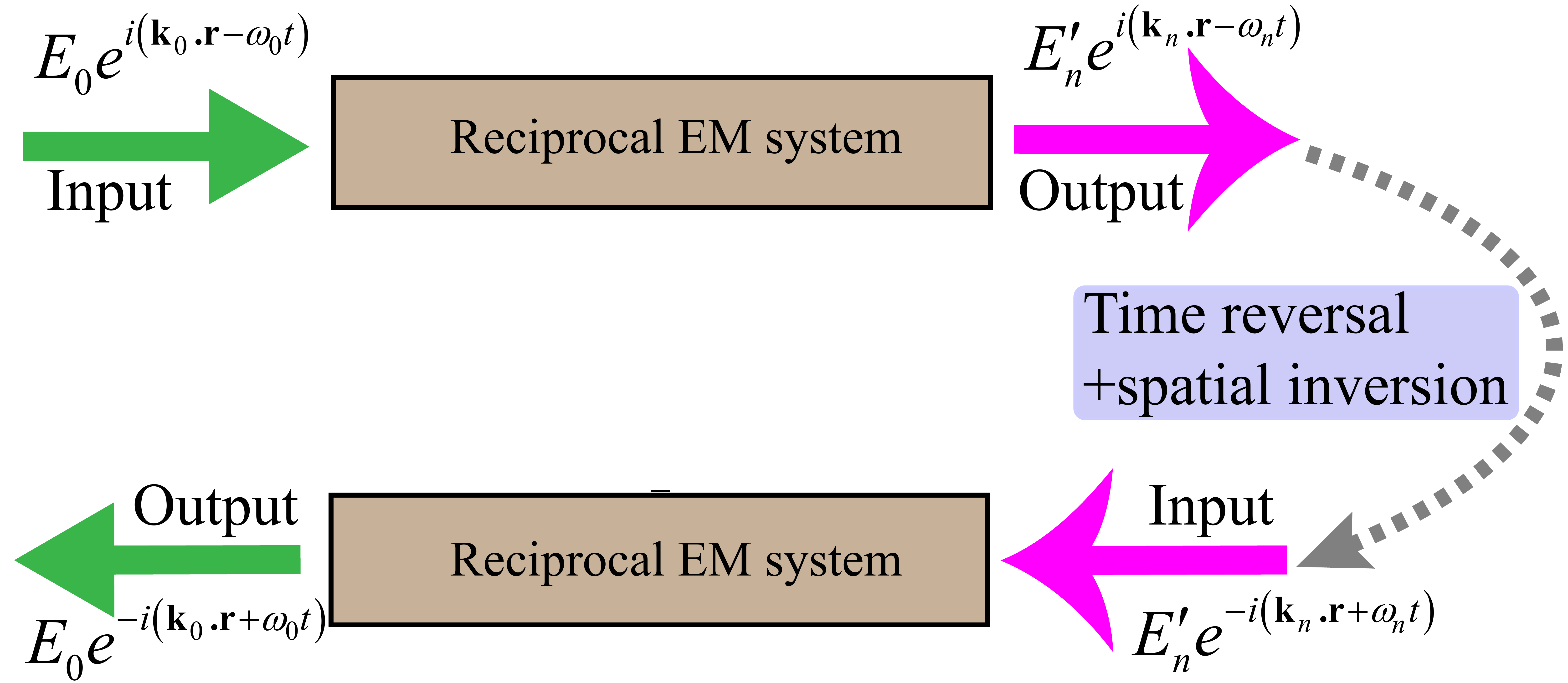}}   
	\caption{Schematic of the experimental set-up configurations for validation of symmetric and reciprocal response of electromagnetic systems. (a) The electromagnetic symmetry of the system is validated, in which the backward problem is the \textit{spatial inversion} of the forward problem. (b) The electromagnetic reciprocity of the system is validated, in which the backward problem is the \textit{spatial inversion of the time-reversed} forward problem.} 
	\label{fig:222}
\end{figure}

The difference between the excitation and response for validation of the symmetry and reciprocity of electromagnetic systems, associated with new frequency generation, is clarified in Figs.~\ref{Fig:asym} and~\ref{Fig:nonrec}. Figure~\ref{Fig:asym} shows the forward and backward problems for the symmetry test of a particular symmetric electromagnetic system, where the backward problem is represented by the spatial inversion of the forward problem, i.e., the applied excitation wave (input) of the backward problem must be the spatial inversion of the excitation wave (input) of the forward problem. As a result, for a symmetric system, the output of the backward problem would be exactly the spatial inversion of the output of the forward problem. Otherwise, the system is asymmetric. Figure~\ref{Fig:nonrec} shows the forward and backward problems for the reciprocity test of a particular reciprocal electromagnetic system, where the backward problem is the spatial inversion of the \textit{time-reversed} of the forward problem, i.e., the applied excitation wave (input) of the backward problem must be the spatial inversion of the output of the forward problem. As a result, for a reciprocal system, the output of the backward problem would be exactly the spatial inversion of the input of the forward problem. Otherwise, the system is nonreciprocal.

Let us now elaborate on the nonreciprocal, asymmetric and angle-asymmetric transmission and reflection of spatial-temporal diffractions introduced by STP gratings that can be used for the realization of a class of efficient telecommunication and optical systems.

\subsubsection{Transmissive STP Grating}
\begin{figure*}
			\includegraphics[width=1.8\columnwidth]{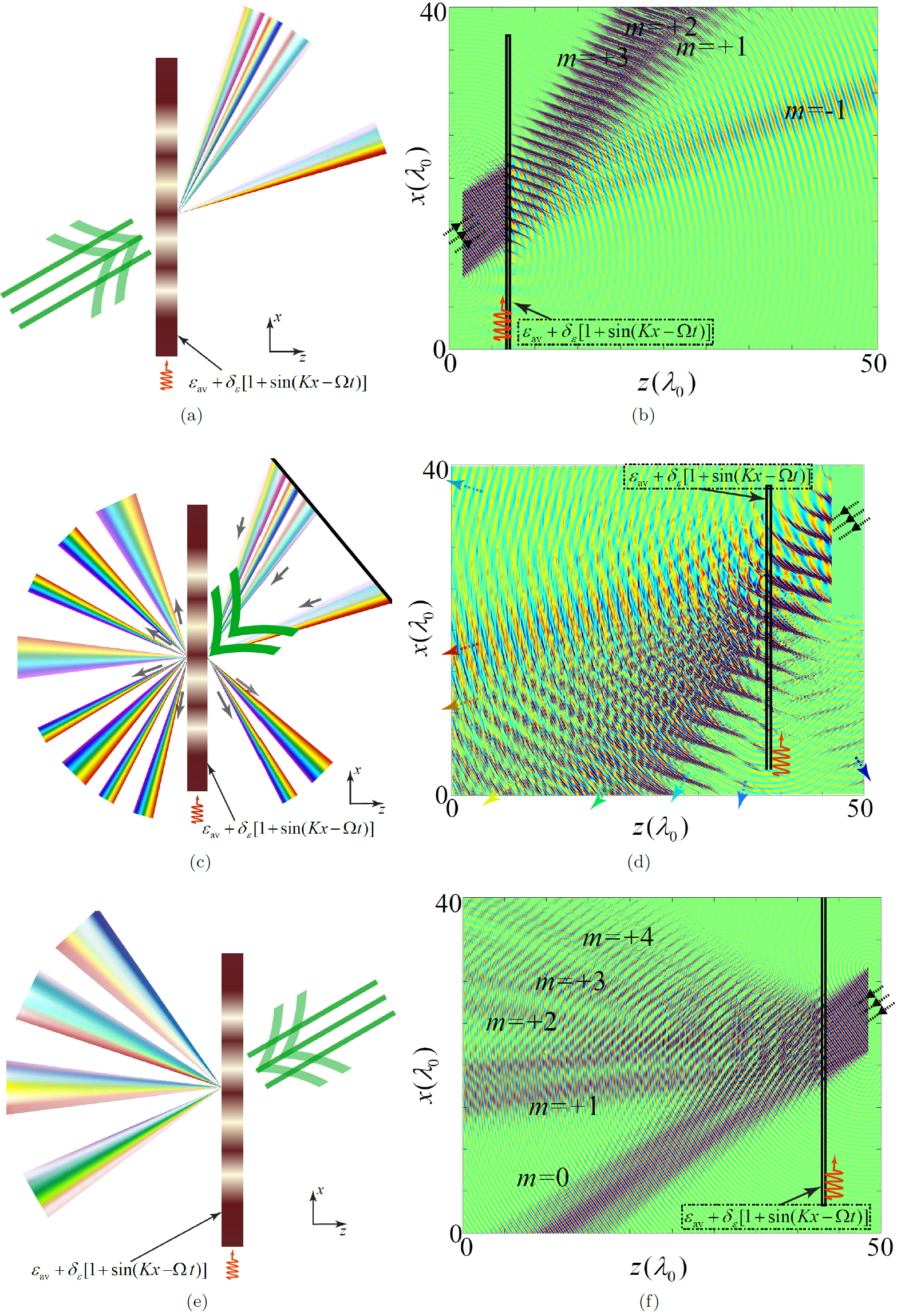}
		\caption{Nonreciprocal and asymmetric wave diffraction from a transmissive STP grating with sinusoidal space-time-varying permittivity, i.e., $\epsilon(x,t)=\epsilon_\text{av}+\delta_\epsilon \sin(K x-\Omega t)$, where $\theta_\text{i}=35^\circ$, $\omega_0=2\pi\times10$~GHz, $\delta_{\epsilon}=0.5$, $\Omega=2\pi\times4$~GHz and $d=0.8 \lambda_0$. (a) and (b) Forward problem. (c) and (d) Backward problem for demonstration of nonreciprocal wave diffraction. (e) and (f) Backward problem for demonstration of asymmetric wave diffraction.}  
		\label{Fig:tr_nr_}
\end{figure*}

Figure~\ref{Fig:tr_nr_}(a) illustrates a particular example, where a $+z$-propagating incident field (forward problem) obliquely impinges on a STP grating. The STP grating possesses an $x$-traveling space-time-varying permittivity $\epsilon(x,t)=\epsilon_\text{av}+\delta_\epsilon [1+\sin(K x-\Omega t)]$. Figure~\ref{Fig:tr_nr_}(b) shows the FDTD numerical simulation result of the transmissive diffraction by the STP diffraction grating in Fig.~\ref{Fig:tr_nr_}(a) with $\theta_\text{i}=35^\circ$, $\omega_0=2\pi\times10$~GHz, where $\delta_{\epsilon}=0.5$, $\Omega=2\pi\times4$~GHz, $d=0.8 \lambda_0$. As expected, the diffracted spatial-temporal orders possess different wavelengths and different amplitudes. Next, we investigate the nonreciprocity of the STP grating in Fig.~\ref{Fig:tr_nr_}(b). Figures~\ref{Fig:tr_nr_}(c) and~\ref{Fig:tr_nr_}(d) show, respectively, the schematic and results of the backward problem for the reciprocity test of the grating. Following the procedure for the reciprocity test shown in Fig.~\ref{Fig:nonrec}, the backward problem is the spatial inversion of the time-reversed of the forward problem in Figs.~\ref{Fig:tr_nr_}(a). Therefore, the excitation input wave of the backward problem is the spatial inversion of the output of the forward problem, which is a polychromatic wave. As it is shown in Figs.~\ref{Fig:tr_nr_}(c) and~\ref{Fig:tr_nr_}(d), the output of the backward problem is completely different than (the spatial inversion of) the incident wave of the forward problem. Hence, the STP grating introduces nonreciprocal wave transmission.

Then, we investigate symmetrical diffraction transmission by the STP grating. Consider a $-z$-propagating incident field (backward wave incidence) that obliquely impinges on the same STP grating as in Figs~\ref{Fig:tr_nr_}(a) and~\ref{Fig:tr_nr_}(b), but from the other side of the STP grating and under the angle of incidence $\theta_\text{i}=35^\circ$. This scenario is depicted in Fig.~\ref{Fig:tr_nr_}(e), and the corresponding time domain response is shown in Fig.~\ref{Fig:tr_nr_}(f). Comparing the numerical simulation results in Figs.~\ref{Fig:tr_nr_}(b) and~\ref{Fig:tr_nr_}(f), we see that the STP grating introduces completely different diffraction patterns for forward and backward incidence. This includes, difference in the angle of diffraction and amplitude of the diffracted fields.

\subsubsection{Reflective STP Grating}
\begin{figure*}
		\includegraphics[width=2\columnwidth]{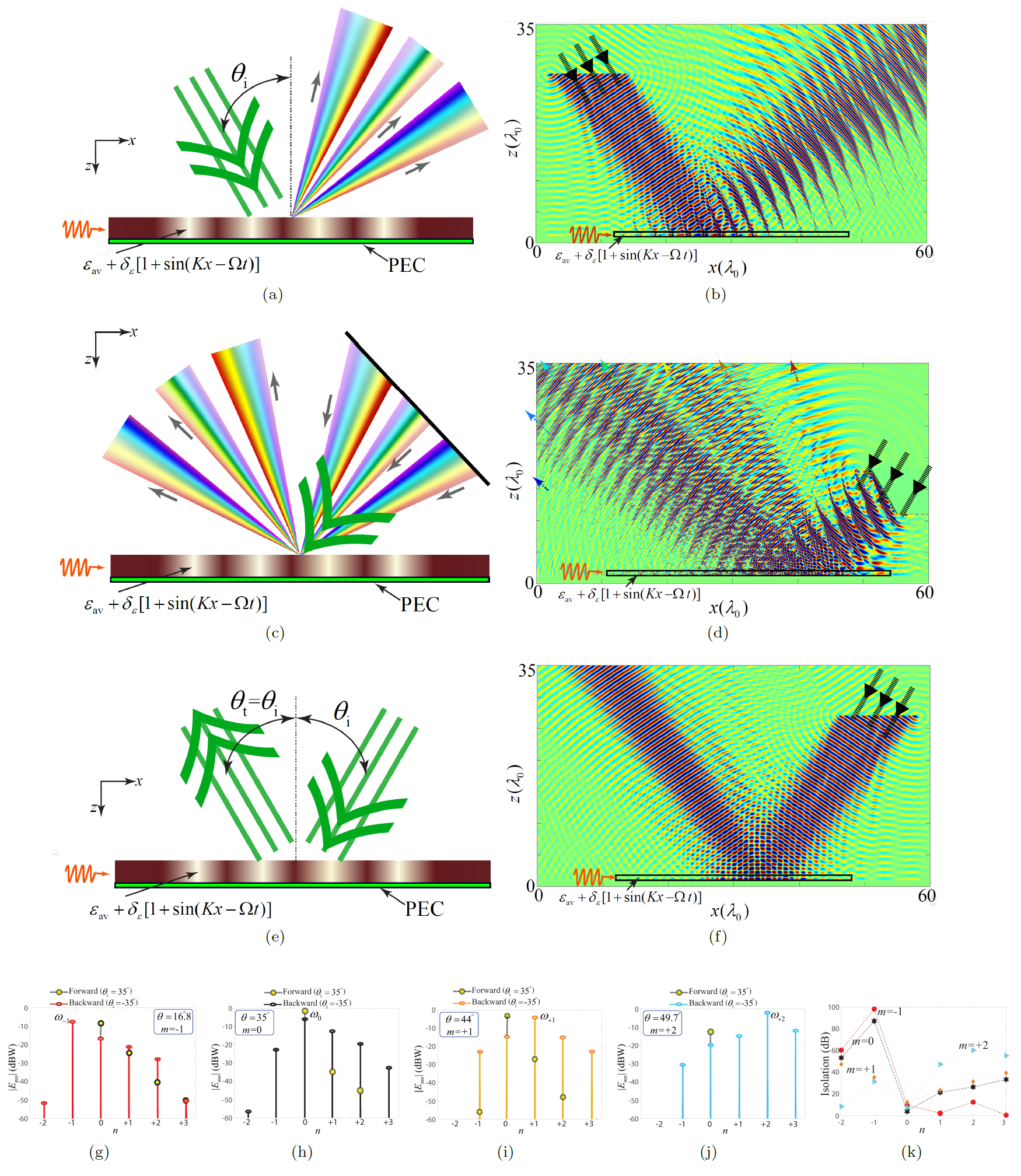}
	\caption{Nonreciprocal and angle-asymmetric spatial-temporal diffraction of a reflective STP diffraction grating with a $+x$-traveling space-time-varying electric permittivity, i.e., $\epsilon(x,t)=\epsilon_\text{av}+\delta_\epsilon [1+\sin(K x-\Omega t)]$, where $\delta_{\epsilon}=0.5$, $\Omega=0.4 \omega_0$, $d=0.8 \lambda_0$. (a) and (b) Forward incidence, where $\theta_\text{i}=35^\circ$. (c) and (d) Backward problem for demonstration of nonreciprocal wave diffraction. (e) and (f) Backward problem for demonstration of angle-asymmetric wave diffraction, where $\theta_\text{i}=-35^\circ$. (g)-(j) Frequency spectrum of the diffracted orders for forward ($\theta_\text{i}=35^\circ$) and backward problem in (e) and (f) ($\theta_\text{i}=-35^\circ$), exhibiting high isolation between differaction orders of forward and backward problems. (k) Isolation between forward and backward diffracted orders achieved from the results in (g)-(j). The analytical results for the angle of diffraction are listed in Table~\ref{tab:table2}.}  
	\label{Fig:a}
\end{figure*}

Here, we study the operation of the STP diffraction grating in the reflective mode. Such a grating may be realized based on the combination of a STP diffraction grating and a metallic surface. Metals naturally reflect light with high efficiency, so that by integrating a STP grating and a metal, one may achieve a fully reflective STP diffraction grating. Figure~\ref{Fig:a}(a) depicts a reflective STP grating, where a perfect electric conductor (PEC) is used at the bottom of the structure providing full reflection of spatial-temporal diffractions. Figure~\ref{Fig:a}(b) provides the numerical results for the diffraction by the grating in Fig.~\ref{Fig:a}(a) for forward incidence, where $\theta_\text{i}=35^\circ$, and with an $+x$-traveling modulation, i.e., $\epsilon(x,t)=\epsilon_\text{av}+\delta_\epsilon [1+\sin(K x-\Omega t)]$, with $\omega_0=2\pi\times10$~GHz, where $\delta_{\epsilon}=0.5$, $\Omega=2\pi\times4$~GHz, $d=0.8 \lambda_0$. Following the same operation as the transmissive STP grating, here the diffracted orders possess different wavelengths.

We next investigate the nonreciprocity of the reflective STP grating in Fig.~\ref{Fig:a}(a). Figures~\ref{Fig:a}(c) and~\ref{Fig:a}(d) show, respectively, the schematic and numerical results of the backward problem for reciprocity test of the reflective grating. Following the procedure for the reciprocity test shown in Fig.~\ref{Fig:nonrec}, the backward problem is the spatial inversion of the time-reversed of the forward problem in Fig.~\ref{Fig:a}(a). Thus, the excitation wave of the backward problem is the spatial inversion of the output of the forward problem, which is a polychromatic wave. It may be seen from Fig.~\ref{Fig:a}(c) and~\ref{Fig:a}(d) that the output of the backward problem is totally different than (the spatial inversion of) the incident wave of the forward problem, which demonstrates strong nonreciprocity of the reflective STP grating.

For the angle-symmetry test of the reflective grating in Fig.~\ref{Fig:a}(a), we consider incidence of the wave under the angle of incidence $\theta_\text{i}=-35^\circ$, as sketched in Fig.~\ref{Fig:a}(e). The corresponding FDTD numerical simulation result is shown in Fig.~\ref{Fig:a}(f). Comparing the results of the forward and backward incidence, shown in Figs.~\ref{Fig:a}(b) and~\ref{Fig:a}(f), respectively, one may obviously see that the reflective diffraction by the grating is completely angle-asymmetric. Such an asymmetric reflective diffraction includes asymmetric angles of diffraction and unequal amplitudes of the diffracted orders. Figures~\ref{Fig:a}(g) to~\ref{Fig:a}(j) plot the FDTD numerical simulation frequency domain responses for the $m=-1$ to $m=+2$ diffracted orders. Figure~\ref{Fig:a}(k) plots the isolation of each spatial-temporal diffracted order for forward and backward incidences shown in Figs.~\ref{Fig:a}(b) and~\ref{Fig:a}(f), respectively. Table~\ref{tab:table2} lists the analytical results for the diffraction angles $\theta_{mn}$ (in degrees) of the reflective STP diffraction grating, considering an incident field impinging on the surface of the grating under the incident angle of $\theta_\text{i}=35^\circ$ for forward incidence and $\theta_\text{i}=-35^\circ$ for backward incidence.

 \begin{table}
	\centering
	\caption{Analytical results for diffraction angles $\theta_{mn}$ (in degrees) of the reflective STP diffraction grating, where the FDTD numerical simulation results are given in Fig.~\ref{Fig:a}.} 
	\label{tab:table2}
	\begin{tabular}{|c||c|c|c|c|c|c|c|c|  }
		\hline
	&  & \multicolumn{7}{c}{$m$} \\
		&Incidence&$-3$& $-2$& $-1$& $0$& $+1$ &$+2$&$+3$\\
		\hline 	\hline
		$n=-3$& Forw.:& Ev. &Ev.&-60  &Ev. &Ev. &Ev. &Ev.\\ 
		&Backw.: &Ev.   &Ev.  &Ev.&  Ev. &  - 60.2&  Ev.&Ev. \\ 	\hline		
		$n=-2$ & Forw.:& Ev. &   Ev. &    60&  Ev. &    Ev. & Ev. &  Ev.  \\ 
     	&Backw.: &Ev.   &Ev.  &Ev.&  Ev. &   60.2&  Ev.&Ev.  \\ \hline
		$n=-1$& Forw.:&  Ev. &  -22.2 &   \underline{\textbf{16.8}}  &  72.9 &  Ev. &  Ev. &   Ev. \\ 
		&Backw.: &Ev.   &Ev.  &Ev.&  73&   16.8&  -22.1&Ev.  \\ \hline 
		$n=0$& Forw.: &-38.8 & -13.1 &   10 &  \underline{\textbf{35}} &  76.8 &  Ev. &   Ev.  \\ 
	    &Backw.:	&Ev.&  Ev.&  76.8&  \underline{\textbf{35}}&   10 & -13& -38.8 \\ \hline
		$n=+1$& Forw.: & -26.6 &  -9.3 &   7.1 & 24.2& \underline{\textbf{44}} &  78.8 & Ev.  \\
		&Backw.: &   Ev.&  78.8&  44&  24.2&  \underline{\textbf{7.1}}&  -9.3& -26.6 \\ \hline
		$n=+2$&  Forw.: &-20.4 & -7.2 &  5.5 & 18.6 & 32.7 & \underline{\textbf{49.7}} & 80.2   \\
		&Backw.: &	 80.2&   49.7 &  32.7&   18.6&    5.5& \underline{\textbf{-7.2}}& -20.4\\ \hline
		$n=+3$& Forw.:&-16.5  & -5.9&   4.5  & 15.1 &  26.3 &  38.6&   \underline{\textbf{53.7}} \\
		&Backw.: &53.7  & 38.6&   26.3&   15.1  &  4.5  & -5.9  &\underline{\textbf{-16.5}}
		 \\\hline
	\end{tabular}
\end{table}	

\section{Application of STP Gratings to Modern Wireless Communication Systems}\label{sec:app}
The proposed STP grating offers unique properties that can be utilized for the realization of new types of electromagnetic devices and operations, such as for instance, nonreciprocal beam shaping and beam coding, multi-functionality antennas, tunable and nonreciprocal beam steering, enhanced resolution holography, multiple images holography, illusion cloaking, etc.

\begin{figure*}
	\includegraphics[width=2\columnwidth]{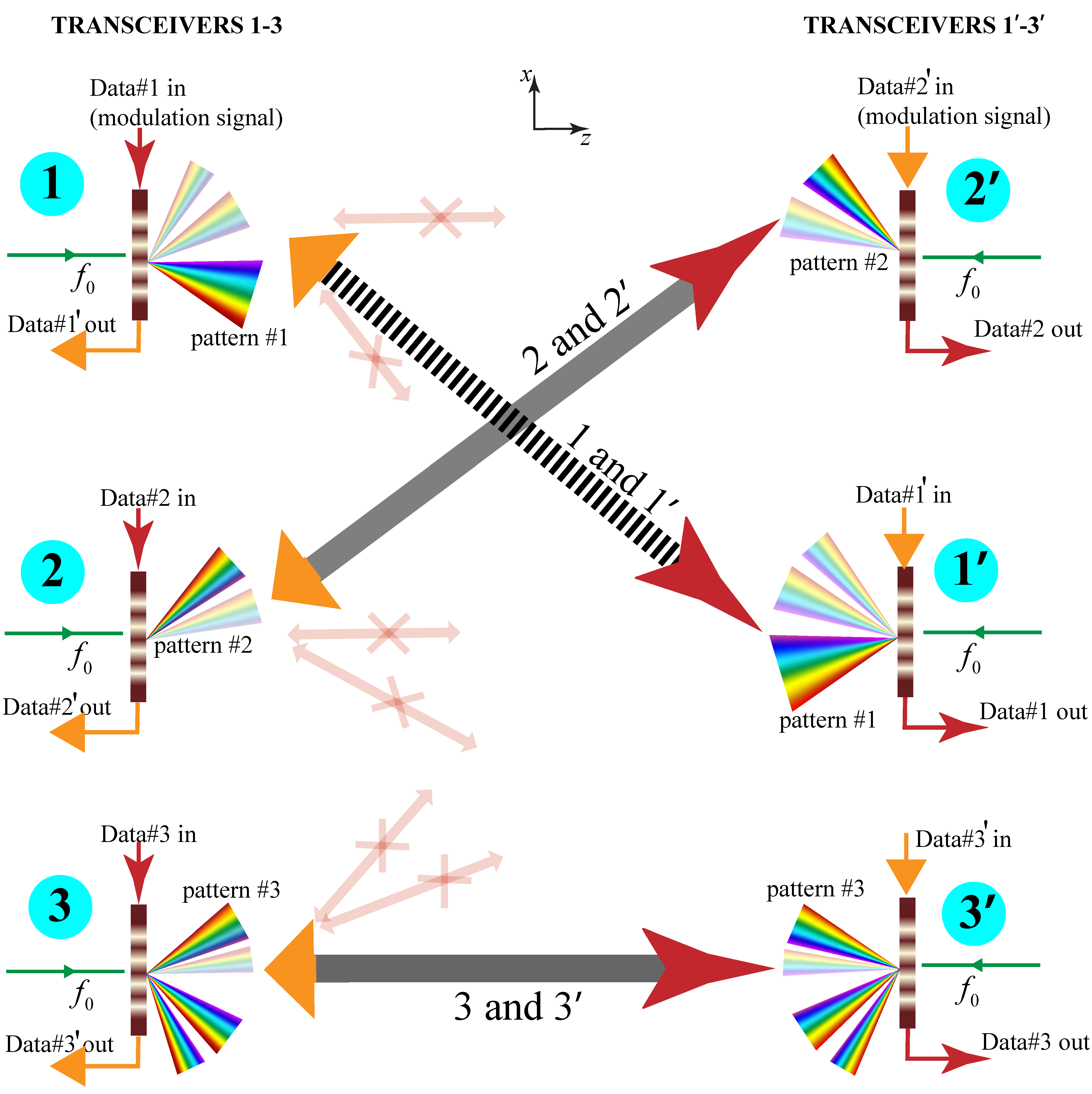}
	\caption{Application of STP diffraction gratings to a full-duplex space-time diffraction code multiple access (STDCMA) system.}
			\label{fig:STDCMAS}
\end{figure*}

Figure~\ref{fig:STDCMAS} presents an original application of the STP diffraction grating to wireless communications. Such a communication system is hereby called space-time diffraction code multiple access (STDCMA) system. In the example provided in Fig.~\ref{fig:STDCMAS}, we consider three pairs of transceivers (in practice one may consider more pairs of transceivers). In such a scenario, only the transceiver pairs that share the same space-time diffraction pattern can communicate. Each diffraction pattern is attributed to the properties of the grating space-time modulation, i.e., the input frequency, where the input data (message) plays the role of the modulation signal. For a specified input data (modulation signal), a unique diffraction pattern is created. In the particular example in Fig.~\ref{fig:STDCMAS}, the transceiver pairs that are allowed to communicate are $1$ and $1'$, $2$ and $2'$, and $3$ and $3'$, so that the transceivers $2'$ and $3'$ ($2$ and $3$) are incapable of retrieving the data sent by the transceiver $1$ ($1'$), and the transceivers $1'$ and $3'$ ($1$ and $3$) are incapable of retrieving the data sent by the transceiver $2$ ($2'$), and so forth. Each communication pair shares a certain space-time diffraction \textit{pattern}. Each diffraction pattern can be created by certain space-time modulation parameters, e.g. $\delta_{\epsilon}$, $\epsilon_\text{av}$, the $K/\Omega$ ratio, and the grating thickness $d$. Since the radiation pattern provided by a STP diffraction grating is very diverse and is very sensitive to the space-time modulation parameters, an optimal isolation between the transceivers can be achieved by proper design of the diffraction patterns.

\begin{figure}
	\includegraphics[width=1\columnwidth]{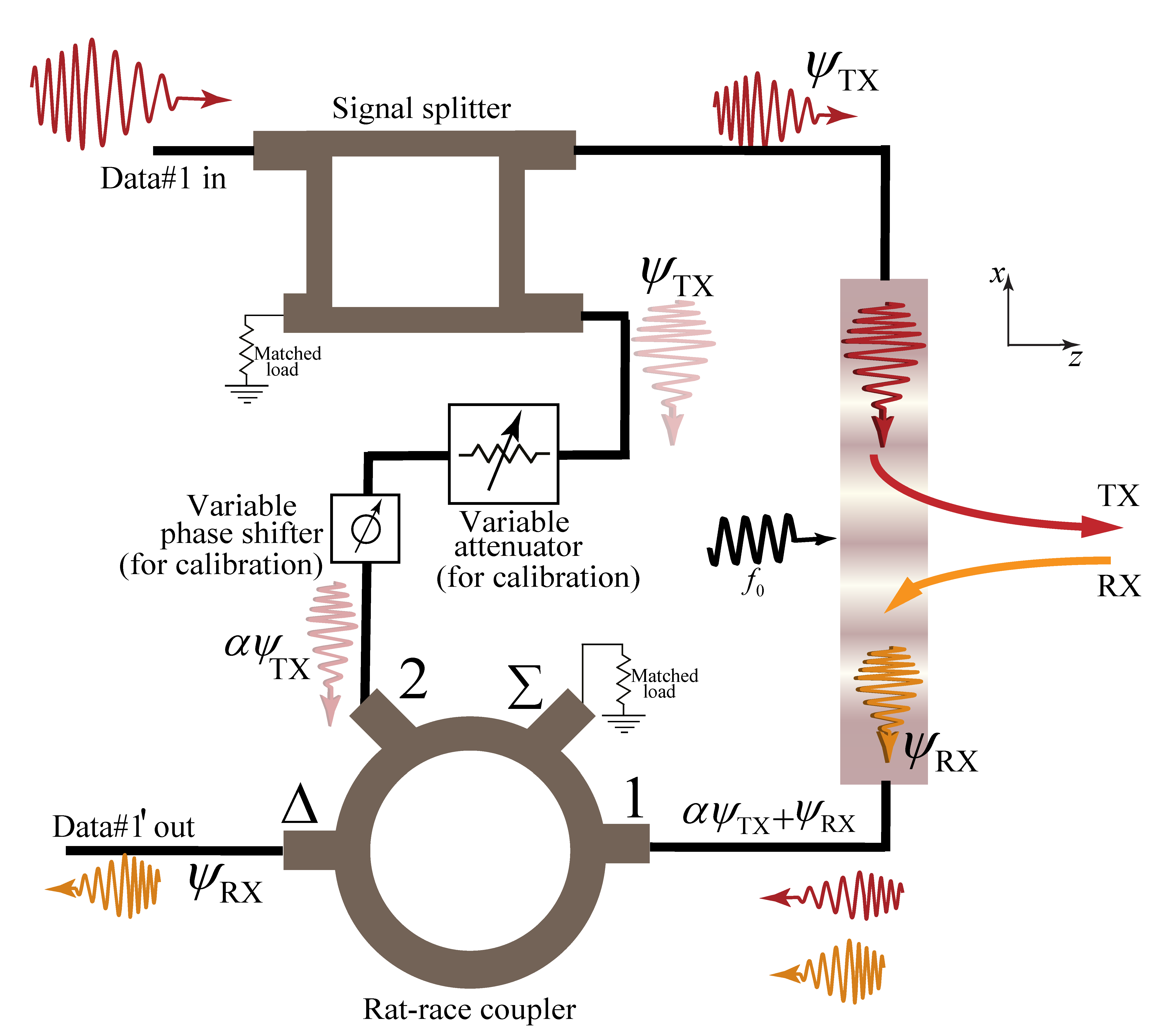}
	\caption{Full-duplex operation of STP-diffraction-based transceivers in Fig.~\ref{fig:STDCMAS} based on unidirectional ($-x$ direction) traveling of both TX and RX signals, where complete cancellation of the TX signal in the RX port is enforced.}
	\label{fig:STDCMAS_circ}
\end{figure}

Such a multiple access scheme is endowed with full-duplex operation, thanks to the unique nonreciprocity provided by the properties of a STP diffraction grating. Figure~\ref{fig:STDCMAS_circ} depicts the architecture of a STP-diffraction-grating-based transceiver in the STDCMA system in Fig.~\ref{fig:STDCMAS}. Such an architecture is composed of a STP diffraction grating illuminated by an incident wave with frequency $f_0$. In the transmit mode (TX), the grating is modulated by the input data denoted by $\psi_\text{TX}$ which is injected to the grating from the top and travels in the $-x$ direction. In the receive mode (RX), the incoming wave (which includes a set of spatial-temporal diffraction orders) impinges on the grating and while interacting with the incident wave with frequency $f_0$, yields a $-x$ traveling wave inside the grating, denoted by $\psi_\text{RX}$. We shall stress that traveling of the $\psi_\text{RX}$ signal in the $-x$ direction is enforced by proper design of the grating, which will be explained later.

As we see in Fig.~\ref{fig:STDCMAS_circ}, the signal wave at the receiver port is composed of the received signal ($\psi_\text{RX}$) plus a portion of the input data of the transmission mode ($\alpha \psi_\text{TX}$). To ensure complete cancellation of the $\psi_\text{TX}$ in the receiver port, we may use the circuit in the left side of Fig.~\ref{fig:STDCMAS_circ}. This circuit is composed of a signal splitter that provides a sample from the input data of the transmit mode ($\psi_\text{TX}$), a variable attenuator and a variable phase shifter for calibration purposes to provide $\alpha \psi_\text{TX}$. Then, the signal wave at the receiver port, i.e., $\psi_\text{RX}+\alpha \psi_\text{TX}$, will be subtracted from the calibrated sample signal, that is $\alpha \psi_\text{TX}$, by a rat-race coupler. Thus, the signal at the difference port of the rat-race coupler is the desired received signal $\psi_\text{RX}$. It is worth mentioning that the calibration of the architecture can be performed by disconnecting the RX port from port-1 of the rat-race coupler, connecting port-1 to a match load, and then seeking for a null at the difference port of the rat-race coupler by adjusting the variable attenuator and variable phase shifter, so that $\psi_\text{TX}$ is completely canceled out at the difference port of the rat-race coupler.
\begin{figure*}
	\includegraphics[width=2\columnwidth]{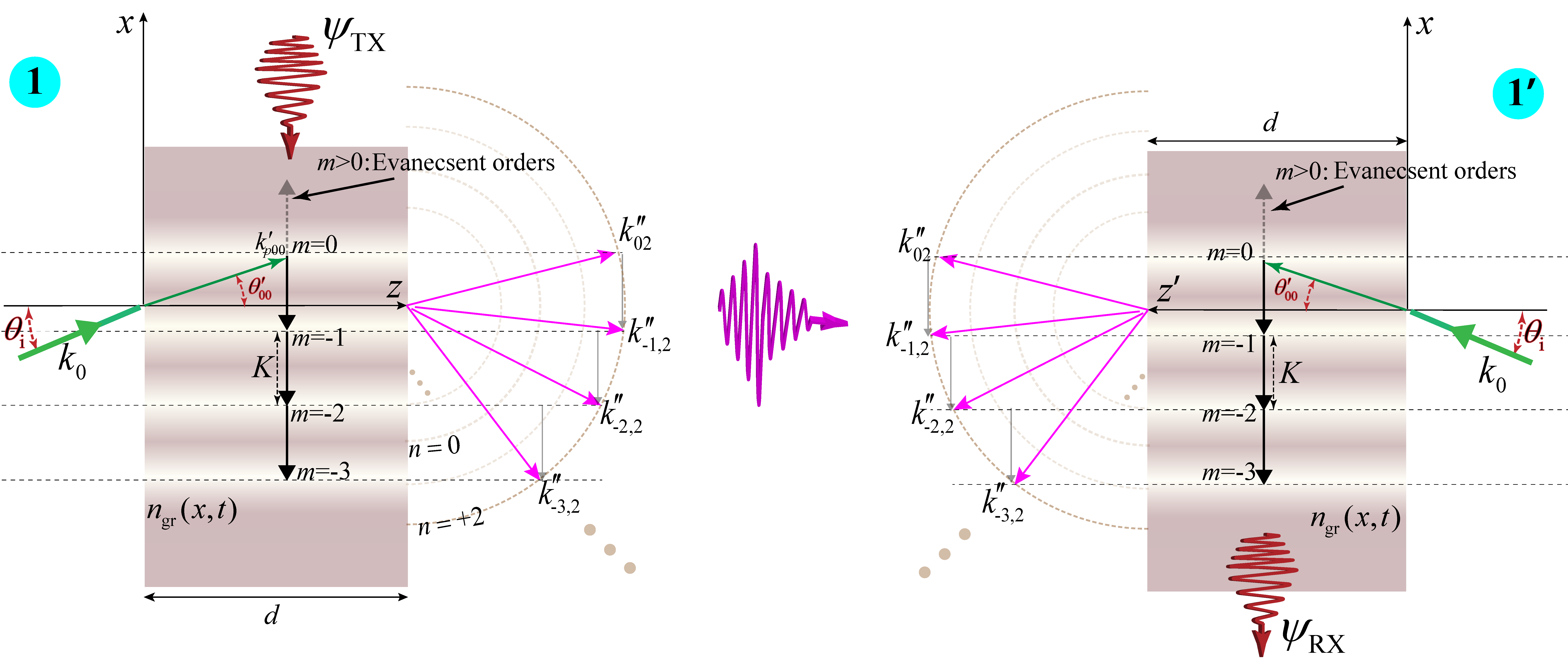}
	\caption{Illustrative wavevector diagram of a particular transceiver pair in the space-time diffraction code multiple access system in Fig.~\ref{fig:STDCMAS}.}
	\label{fig:STDCMAS_wv}
\end{figure*}

An elegant feature of the transceiver scheme in Fig.~\ref{fig:STDCMAS_circ} is that the diffraction grating is used as the receiver (as well as the transmitter), where the received signal wave acts as the modulation signal (specifies the $K$ and $\Omega$ parameters) instead of the incident field. The key reason for the duplexing operation is that, inside the grating the wave can only flow downstream. Figure~\ref{fig:STDCMAS_wv} shows how the full-duplex operation is achieved by proper design of the diffraction grating, where only negative diffraction orders, i.e., $-x$ propagating orders, are generated, while positive diffraction orders, that is $+x$ propagating orders, are evanescent. This way, we ensure that inside the grating, all the diffraction orders are traveling in the $-x$ direction, in both the transmit and receive modes. In Fig.~\ref{fig:STDCMAS_wv}, transceiver $1$ operates in the transmit mode, where the input data ($\psi_\text{TX}$) is injected to the grating from the top and while interacting with the incident wave with the wavenumber $k_0$, generates a number of nonpositive diffraction orders, i.e., $m \geq 0$. In the right side of Fig.~\ref{fig:STDCMAS_wv}, transceiver $1'$ receives the diffracted orders by transceiver $1$, so that the resultant wave inside $\psi_\text{RX}$ exits the grating from the bottom port of the grating, as all the diffraction orders can only travel in the $-x$ direction.
\vspace{6mm}
\section{Conclusion}\label{sec:conc}
We have presented the analysis and characterization of space-time periodic (STP) diffraction gratings as the generalized version of conventional spatially periodic diffraction gratings. Such STP gratings offer enhanced functionalities and exotic behaviour. It is shown that such gratings provide an asymmetric diffraction pattern, nonreciprocal diffraction, and an enhanced diffraction efficiency as well as frequency generation. Moreover, each spatial diffraction order includes an infinite set of temporal diffraction orders. We provided the theoretical investigation of the problem, which has been supported by FDTD numerical simulation results. Such structures with marked differences with conventional spatially periodic diffraction gratings are expected to find interesting applications in optical and communication systems. As a particular example, we have proposed the space-time diffraction code multiple access (STDCMA) system, as a promising communication system featuring full-duplex operation.

\bibliographystyle{ieeetr}
\bibliography{Taravati_Reference}

\end{document}